\documentclass[a4paper, amsfonts, amssymb, amsmath, reprint, prl, showkeys, nofootinbib, twoside, superscriptaddress, longbibliography]{revtex4-2}

\usepackage{amsmath}
\usepackage{amsfonts}
\usepackage{amssymb}
\usepackage{placeins}
\usepackage{changes}
\usepackage[english]{babel}
\usepackage[utf8]{inputenc}
\usepackage{parskip}
\usepackage{changes}
\usepackage[pdftex, pdftitle={Article}, pdfauthor={Author}]{hyperref} % For hyperlinks in the PDF

\newcommand{\Rnl}{\ensuremath{R_\mathrm{nl}}}
\newcommand{\Rmax}{\ensuremath{R_\mathrm{max}}}

\newcommand{\Vbg}{\ensuremath{V_\mathrm{bg}}}
\newcommand{\Vtg}{\ensuremath{V_\mathrm{tg}}}

\newcommand{\DBmax}{\ensuremath{\Delta B_\mathrm{max}}}

\begin{document}
\title{A ballistic electron source with magnetically-controlled valley polarization in bilayer graphene}

\author{Josep Ingla-Ayn\'es}
    \email[ ]{jingla@mit.edu}% Your name
    \affiliation{Kavli Institute of Nanoscience, Delft University of Technology, Lorentzweg 1, 2628 CJ Delft, The Netherlands}
\author{Antonio L. R. Manesco}
    \affiliation{Kavli Institute of Nanoscience, Delft University of Technology, Lorentzweg 1, 2628 CJ Delft, The Netherlands}
\author{Talieh S. Ghiasi}
    \affiliation{Kavli Institute of Nanoscience, Delft University of Technology, Lorentzweg 1, 2628 CJ Delft, The Netherlands}
\author{Kenji Watanabe}
    \affiliation{Research Center for Electronic and Optical Materials, National Institute for Materials Science, 1-1 Namiki, Tsukuba 305-0044, Japan}
\author{Takashi Taniguchi}
    \affiliation{Research Center for Materials Nanoarchitectonics, National Institute for Materials Science,  1-1 Namiki, Tsukuba 305-0044, Japan}
\author{Herre S. J. van der Zant}
    \affiliation{Kavli Institute of Nanoscience, Delft University of Technology, Lorentzweg 1, 2628 CJ Delft, The Netherlands}
\date{\today} % Leave empty to omit a date

\begin{abstract}
The achievement of valley-polarized electron currents is a cornerstone for the realization of valleytronic devices.
Here, we report on ballistic coherent transport experiments where two opposite quantum point contacts (QPCs) are defined by electrostatic gating in a bilayer graphene (BLG) channel. By steering the ballistic currents with an out-of-plane magnetic field we observe two current jets, a consequence of valley-dependent trigonal warping. Tuning the BLG carrier density and number of QPC modes ($m$) with a gate voltage we find that the two jets are present for $m=1$ and up to $m=6$, indicating the robustness of the effect. Semiclassical simulations which account for size quantization and trigonal warping of the Fermi surface quantitatively reproduce our data without fitting parameters, confirming the origin of the signals. In addition, our model shows that the ballistic currents collected for non-zero magnetic fields are valley-polarized independently of $m$, but their polarization depends on the magnetic field sign, envisioning such devices as ballistic current sources with tuneable valley-polarization.

\end{abstract}

\keywords{Valleytronics, bilayer graphene, quantum point contacts, trigonal warping}

\maketitle
\begin{figure}[htb]
	\centering
		\includegraphics[width=0.45\textwidth]{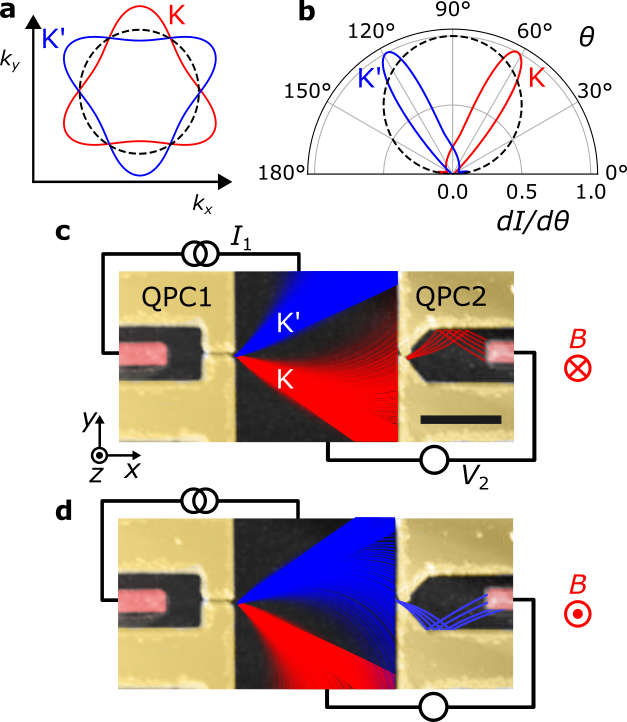}
	\caption{\textbf{Trigonal warping, valley splitter, and magnetically-controlled valley-polarized current source in bilayer graphene (BLG).} (a) Fermi surface of BLG in valley K (red) and K' (blue). A circular Fermi surface (black dashed line) is shown for comparison. (b) Angular distribution  ($dI/d\theta$) of the valley-polarized current jets emitted by a quantum point contact (QPC) in BLG, showing that QPC1 is a valley splitter. The dashed black line corresponds to the conventional case where the distribution is $\cos(\theta)$. (c) False color atomic force microscopy image of the measured device where the split gates, that define the QPCs by preventing electron transport in the covered areas, are yellow and the contacts to BLG are light red. The scale bar is 2~$\mu$m. The red and blue lines represent valley-polarized electron trajectories calculated under a small negative magnetic field ($B$). Reversing $B$ results in the trajectories shown in (d). In panel c, only electrons in valley K reach the detector island. In contrast, in panel d, only K' electrons are collected, illustrating the $B$-controlled valley filter operation.
 }
	\label{Figure1}
\end{figure}

A functional valleytronic device requires controllable injection of valley-polarized currents \cite{schaibley2016,yu2020}. Thus, ballistic valley splitters, which enable current sources with magnetic-field tuneable valley polarization, are crucial components of valleytronic devices. 
For this purpose, Bernal stacked bilayer graphene (BLG) is a unique material \cite{martin2008,san2009,li2016,li2018,yin2022} thanks to its gate-tuneable bandgap \cite{castro2007,oostinga2008,zhang2009,icking2022} and trigonal warping, i.e.~triangular distortion of its Fermi surface, which has opposite sign in each valley \cite{mccann2013} (see Fig.~\ref{Figure1}a). Such a band structure makes quantum point contacts (QPCs) emit valley-polarized ballistic electron jets \cite{garcia2008,wang2010, mccann2013, stegmann2018,manesco2023}. Despite early theoretical proposals, it is only recently that the current jets have been mapped by scanning gate measurements \cite{gold2021} in an electrostatically-defined QPC in BLG \cite{1overweg2018,1kraft2018,overweg2018,kraft2018,knothe2018,lane2019,banszerus2020,lee2020,gall2022}. This is due to the stringent requirements of coherent injection by a QPC with significant trigonal warping \cite{gold2021,manesco2023}. 

\begin{figure*}[!htb]
	\centering
		\includegraphics[width=0.9\textwidth]{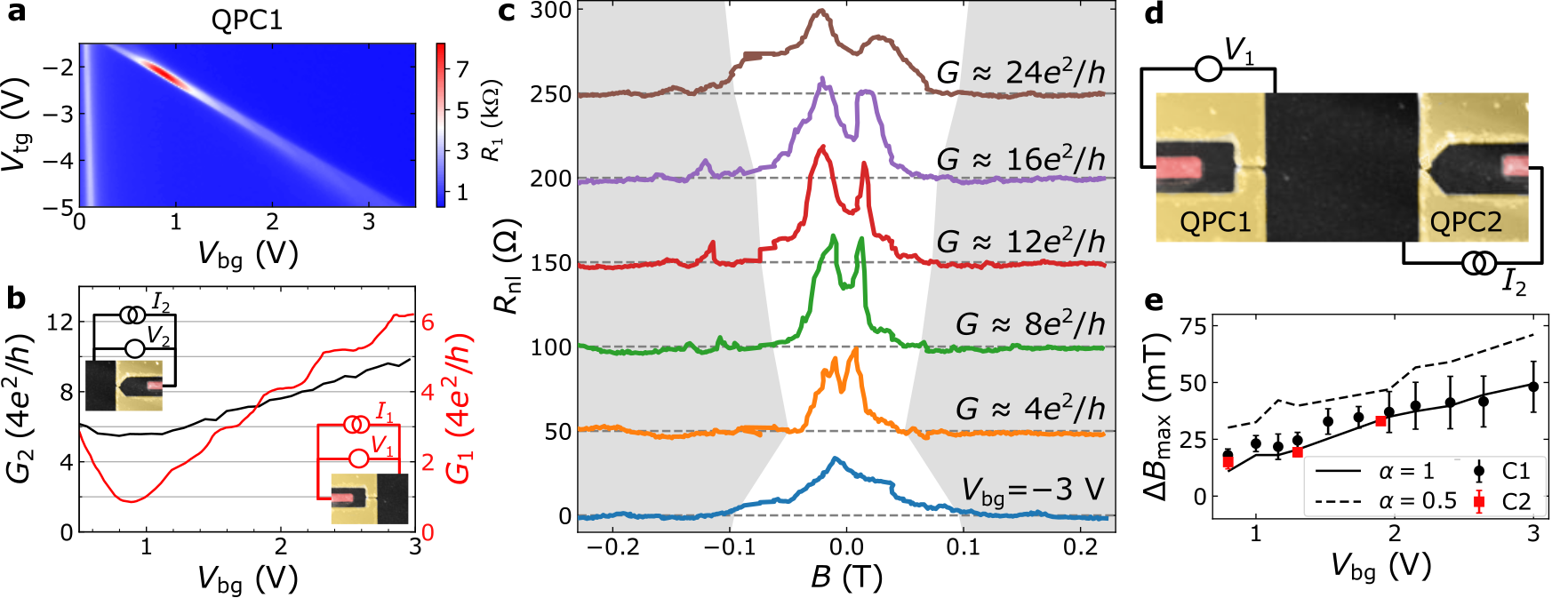}
	\caption{\textbf{Steering of current jets emitted by a QPC using an out-of-plane magnetic field.} (a) Two-terminal resistance measured across QPC1 (right inset in panel b) as a function of the back-gate (\Vbg{}) and split-gate (\Vtg{}) voltages. (b) Conductance of QPC1 (red line, right axis and right inset circuit) and QPC2 (black line, left axis and left inset circuit). (c) Collimation experiments at different \Vbg{}. The positive \Vbg{} values correspond to quantized $G$ and are labeled accordingly. The gray areas mark the $B$-range where the average diameter of the cyclotron orbit is smaller than the QPC separation and no signal is expected. The curves are offset 50~$\Omega$, as shown by the dashed lines. An offset in $B$ was added to correct for the magnet remanence. (d) Measurement geometry corresponding to the collimation experiments in panel c. (e) Peak separation \DBmax{} vs.~\Vbg{}. The black circles are extracted from panel c; the red squares are obtained from measurements on a second pair of QPCs (C2). The black solid (dashed) line is from the model described in the main text and in Fig.~\ref{Figure3} for $\alpha=1$ ($\alpha=0.5$). The error bars are estimated from the peak widths.}
	\label{Figure2}
\end{figure*}

Recent transverse electron focusing experiments \cite{ingla2023} between gate-defined QPCs in BLG showed signatures of valley-resolved ballistic transport. However, as they require a significant electron deflection by the magnetic field, a large fraction of the Fermi surface is probed, and most of the trigonal warping effects are averaged out.
In this context, a ballistic measurement setup where electrons are deflected only by small angles \cite{molenkamp1990,yacoby1991,shepard1992,barnard2017} would unambiguously probe the formation of ballistic valley-polarized electron jets (see Fig.~\ref{Figure1}b), opening the way for new valley polarized electron sources with magnetic-field-controlled polarization.

\begin{figure*}[!htb]
	\centering
		\includegraphics[width=0.95\textwidth]{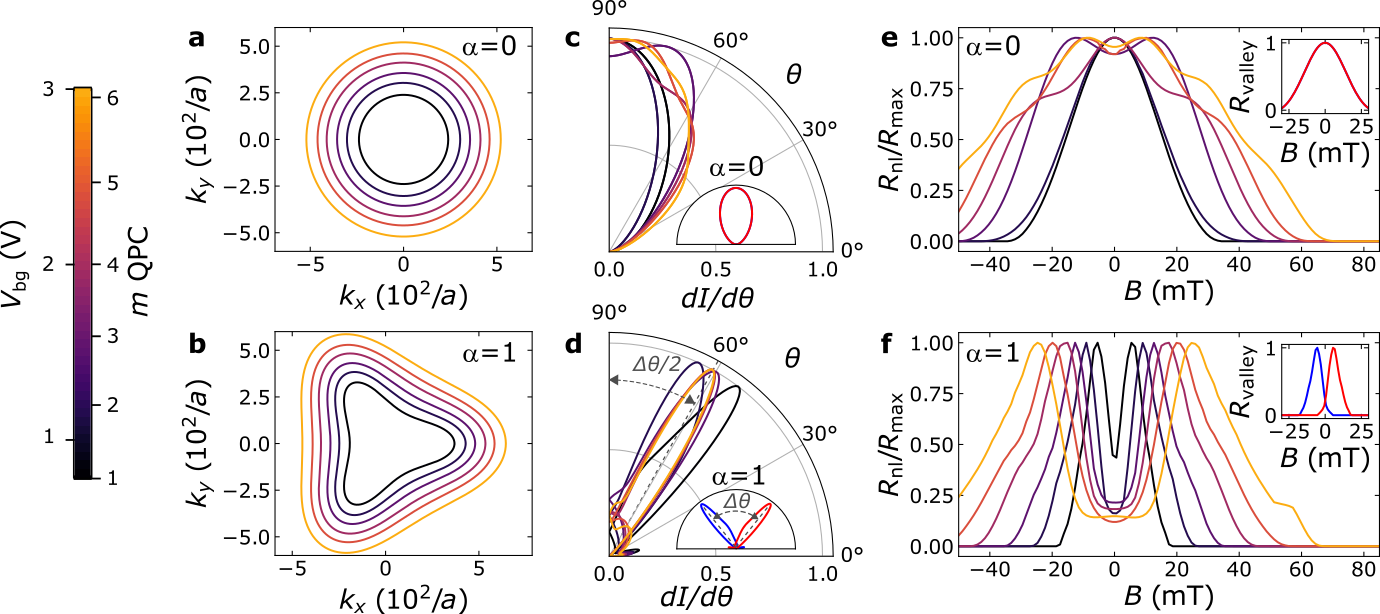}
	\caption{\textbf{Semi-classical simulations and role of trigonal warping at different \Vbg{}.} All the curves (apart from the insets) are color-coded according to the left color bar. The insets are colored according to the valley index (K is red and K' is blue). (a) and (b) show the Fermi surfaces, (c) and (d) show the angular distribution of emitted currents $dI/d\theta$, and (e) and (f) show the simulated collimation spectra without ($\alpha=0$) and with ($\alpha=1$) trigonal warping, respectively. In panel d, $\Delta\theta$ is the angular separation between $dI/d\theta$ peaks. Ignoring the valley polarization, $dI/d\theta$ in panels c and d is symmetric with respect to $\theta=90^\circ$. The insets show the corresponding valley-resolved signal at $\Vbg{}= 0.8$~V, with a single QPC mode. In panels a and b, $a=\sqrt{3}a_0$, where $a_0$ is the in-plane separation between carbon atoms. 
 }
	\label{Figure3}
\end{figure*}

 Here, we perform ballistic collimation experiments where two electrostatically-defined QPCs are placed opposing each other (see Fig.~\ref{Figure1}c), and the valley-polarized current jets are steered toward the detector using an out-of-plane magnetic field ($B$). The measurements show the formation of two ballistic current jets independently of the number of QPC modes ($m$). 
Using semi-classical simulations {that predict the formation of valley-polarized current jets}, we reproduce the experimental features without the need for any fitting parameter. 
These observations confirm that, when current-biased, the measured devices operate like magnetic-field controlled valley-polarized ballistic current sources as illustrated in Figs.~\ref{Figure1}c and \ref{Figure1}d. 

Our valley-polarized electron sources are in a van der Waals heterostructure consisting of a BLG flake encapsulated between an upper and a lower hBN flake with thicknesses 23 and 28~nm, respectively. The stack is placed on a multilayer graphene back gate \cite{1overweg2018}. The device, which is shown in Fig.~\ref{Figure1}c, contains multiple contacts to the BLG flake (red) and 50-nm-separated split gates on the top hBN (yellow); see Ref.~\cite{ingla2023} for fabrication details. 

We first investigate the QPC formation. By applying a bias current ($I_1=100$~nA) to QPC1 and monitoring the two-terminal voltage ($V_1$, right inset of Fig.~\ref{Figure2}b) as a function of the back gate voltage (\Vbg{}) and the voltage applied to the split gates (\Vtg{}), we obtain the resistance map in Fig.~\ref{Figure2}a, where $R_1=V_1/I_1$. 
The vertical line at \Vbg{}$\approx0$ corresponds to the charge neutrality point of the non-top-gated BLG regions and shows that the residual doping is very small. 
The diagonal line corresponds to the charge neutrality point of the double-gated region and gives rise to a maximal $R_1$ at each \Vbg{} [$\Rmax{}(\Vbg{})$]. 
\begin{figure}%[tb]
	\centering
		\includegraphics[width=0.45\textwidth]{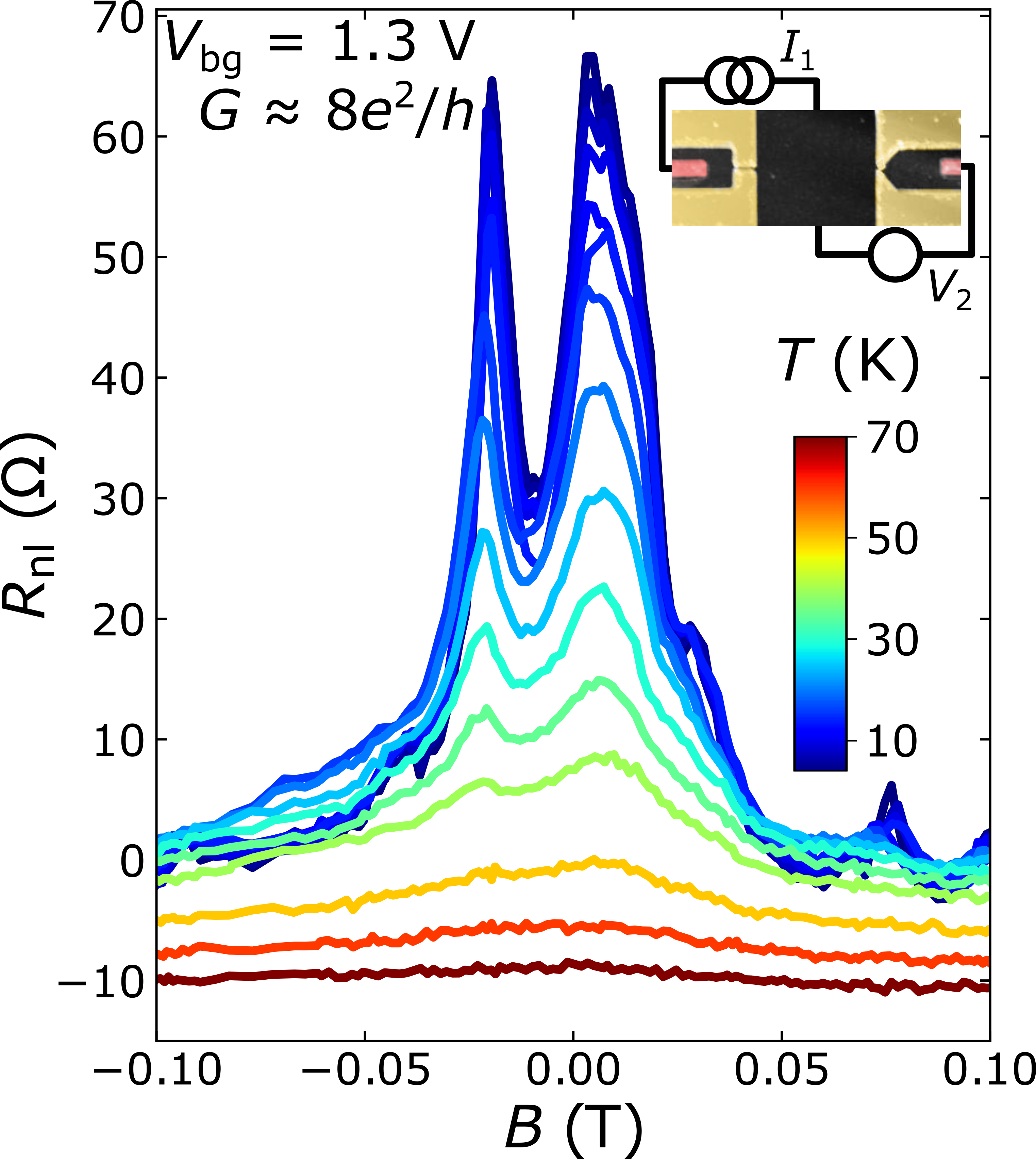}
	\caption{\textbf{Temperature dependence of ballistic electron jetting} at \Vbg{}$=1.3$~V. The plot colors indicate the measurement temperature as illustrated by the color bar. The curves are offset in $B$ to correct for the magnet remanence. The inset shows the measurement geometry, which is the reciprocal of Fig.~\ref{Figure2}. As a consequence, the $B$-asymmetry shows opposite behavior. }
	\label{Figure4}
\end{figure}
For \Vbg{}$>0$, $\Rmax{}(\Vbg{})$ increases with increasing \Vbg{} until \Vbg{}$\approx0.8$~V, after which it starts decreasing. This is due to the formation of a QPC between the split gates with increasing carrier density ($n$). 

To find if the QPC conductance ($G$) is quantized, we determine it using $G(\Vbg{})=(R_\mathrm{max}(\Vbg{})-R_\mathrm{min}(\Vbg{}))^{-1}$, where $R_\mathrm{min}(\Vbg{})$ is the minimal resistance at each \Vbg{} which is dominated by the contact resistances. The result is shown in Fig.~\ref{Figure2}b by the red line and shows plateaus at $G=m\times4e^2/h$, where $m=1$, ..., 6 for \Vbg{}$>0$, indicating the formation of a spin and valley-degenerate QPC with $m$ modes \cite{overweg2018,1kraft2018}.  
As shown in supplementary section 1, for \Vbg{}$<0$ $G\approx8\times 4 e^2/h$ and is not quantized.
To characterize QPC2, we use the black circuit in the left inset of Fig.~\ref{Figure2}b and, applying a current $I_2$, we obtain $R_2=V_2/I_2$ and subtract the contact resistances as above to obtain $G_2$, that is plotted as the drawn black line in Fig.~\ref{Figure2}b. It does not show size quantization steps, neither for electron nor hole transport (supplementary section 1).

To determine the angular distribution of the charge current emitted by the QPCs, we have measured $V_1$ while applying a current $I_2=100$~nA across QPC2 (black circuit in Fig.~\ref{Figure2}d. As shown below, this configuration is equivalent to Fig.~\ref{Figure1}c). The measurement has been performed at different \Vbg{} and, to ensure that charge transport occurs only through the gate-defined QPCs, \Vtg{} has been adjusted to the diagonal line in Fig.~\ref{Figure2}a.
\Rnl{}$=V_1/I_2$ is plotted vs.~$B$ in Fig.~\ref{Figure2}c and, for positive \Vbg{} (quantized $G$), shows two clear peaks for all the measured \Vbg{}, as expected from electron jetting. In contrast, at \Vbg{}$=-3$~V and for all the measured $\Vbg{}<0$ values, even though a structure consistent with two broadened current jets can be distinguished, only a single peak is observed (see supplementary sections 2 and 4 for more details).  

To further understand the measured signals, we quantify the $B$-separation between \Rnl{} peaks (\DBmax{}), which is shown in Fig.~\ref{Figure2}e vs.~\Vbg{} for \Vbg{}$>0$. In this figure, \DBmax{} increases with $n$. Note that, in addition to the data obtained from the QPCs in Fig.~\ref{Figure2}d which we call C1, we have also measured collimation between a second pair of QPCs aligned along the same crystallographic direction, which we call C2 (see supplementary sections 5). \DBmax{} obtained from C2 is shown as the red dots in Fig.~\ref{Figure2}e and shows good agreement with C1.
It is worth noting that, even though the emission of current jets has been reported for a QPC with one mode ($m=1$) \cite{gold2021}, how these will evolve experimentally when higher QPC modes contribute to the angular distribution of emitted currents remains a question. In the absence of trigonal warping, the higher modes give rise to extra lobes in the collimation \cite{shepard1992, topinka2000}, but their effect in the presence of trigonal warping remains unknown to this point. The results in Fig.~\ref{Figure2} show that the peak spacing widens as the number of modes of the detector QPC increases from 1 to 6. The absence of other clear changes indicates that the presence of high-order QPC modes does not prevent current jetting. 

We study the origin of the double-peak feature in $\Rnl{}$ with semiclassical simulations accounting for trigonal warping of the Fermi surface and size quantization at the QPC.
We vary the trigonal warping strength with the parameter $\alpha$: $\alpha=0$ amounts to a circular Fermi surface (Fig.~\ref{Figure3}a), whereas $\alpha=1$ results in the fully warped Fermi surface (Fig.~\ref{Figure3}b) calculated using the parameters obtained in Ref.~\cite{kuzmenko2009}. The model is described in supplementary section 6 and the code used for the simulations is available in Ref.~\cite{zenodo}.

The shape of the Fermi surface influences the angular distribution of injection current $dI/d\theta$.
Because in the absence of trigonal warping the Fermi surface is isotropic, $dI/d\theta$ is equal for both valleys (Fig.~\ref{Figure3}c, inset). 
Since we include size quantization effects on the QPC in our calculations, $dI/d\theta$ varies as a function of the number of modes~\cite{shepard1992,topinka2000}. As illustrated in Fig.~\ref{Figure3}c, for $\alpha=0$ and $m>1$, this results in multiple collimation lobes.
In contrast, for a fully warped Fermi surface (Fig.~\ref{Figure3}d) $dI/d\theta$ becomes valley-dependent and the double-jetting persists regardless of the number of modes, in agreement with \cite{manesco2023}. Note that the exact separation between the $dI/d\theta$ peaks ($\Delta\theta$) is sensitive to \Vbg{}.
This result implies that the current jet formation is sensitive to the electric field and/or carrier density at the QPC. We explain it considering that the Fermi surfaces are not perfect triangles and changing the $k$-dependent occupation of the electron states can influence the current jet orientation.
This result is consistent with our observation that, at fixed \Vbg{}, variations in \Vtg{} (that influence the electric field at the QPC but not the ballistic transport in the black areas of Fig.~\ref{Figure2}d) change the peak positions up to 10~mT at \Vbg{}$=1.3$~V %which corresponds to the uncertainty in Fig.~\ref{Figure2}e 
(see {supplementary section 3}).

Finally, we compute the current absorbed in the collector.
Comparing Figs.~\ref{Figure3}e and \ref{Figure3}f, we find that the double-peak feature occurs for the whole \Vbg{} range only with a warped Fermi surface.
Furthermore, as in the experiment, the peak spacing obtained with a fully warped surface increases with \Vbg{}.
This change in the peak spacing occurs due to the increase of the Fermi surface as the electron density increases (Fig.~\ref{Figure3}b).
We extracted the peak separation from Fig.~\ref{Figure3}f and plotted it vs.~\Vbg{} in Fig.~\ref{Figure2}e as a black solid line.
The agreement between the model with $\alpha=1$ and the experimental results presents compelling evidence that the measured signals are caused by valley-polarized current jets. 
We stress the absence of fitting parameters in this comparison and that all the relevant parameters in the model originate from independent measurements on the same sample and tight-binding parameters obtained by infrared spectroscopy~\cite{kuzmenko2009}. 

It is worth noting that the naive expectation $\DBmax{}\approx2\hbar \sqrt{n\pi}/(eL)$, coming from the assumption of two 60$^\circ$-separated current jets and a circular Fermi surface, overestimates \DBmax{} by approximately a factor of two, stressing the crucial role of trigonal warping on the electron deflection. 
Moreover, we observe that \DBmax{} is overestimated if $\alpha<1$ (dashed line in Fig.~\ref{Figure2}e, and {supplementary section 6D}), demonstrating that \DBmax{} is sensitive to the Fermi surface shape.

The experimental data (see Fig.~\ref{Figure2}c) also shows an asymmetry of the \Rnl{} peak widths. While the spectra calculated in Fig.~\ref{Figure3} assume a perfect alignment of the QPCs with the BLG crystal planes and perfectly opposing QPCs, these assumptions are unlikely to be fulfilled in practice. We have extended the model in supplementary section 6E to account for such misalignments and realized that both can lead to asymmetry in the peaks, in width and height. This result shows that small misalignments can be the reason for the observed peak asymmetry.

We have also investigated the temperature ($T$) dependence of the double-peak structure. Such an experiment has been performed at \Vbg{}$=1.3$~V corresponding to $G=8e^2/h$ and the results are shown in Fig.~\ref{Figure4}. Note that, by swapping the $I$-source and $V$-measurement units with respect to Fig.~\ref{Figure3} (Fig.~\ref{Figure4}, inset), the $B$-asymmetry of \Rnl{} changes sign, confirming that the $I$-source does not affect the current jet formation and our measurements obey reciprocity \cite{buttiker1986, buttiker1988}. Besides a very clear peak width asymmetry, \Rnl{} also shows a slow decay as $T$ increases, leading to the observation of the two peaks up to 40~K. 
Above this temperature, \Rnl{} becomes negative at $|B|\approx0.1$~T. The negative signal measured for $T\geq50$~K can be due to either a significant portion of the current propagating across the double-gated regions or diffusive charge transport between the injector and collector. Furthermore, the observed $T$-dependence of the double-peak near $B=0$ confirms that it is not caused by quantum interference between ballistic trajectories, as the phase coherence length in BLG is expected to decrease quickly with increasing $T$, suppressing interference effects \cite{kozikov2012} above 10~K (see {supplementary section 5} for a more detailed analysis).

To conclude, we have performed collimation experiments between electrostatically-defined QPCs and detected ballistic electron jets. Using semi-classical simulations, we show that the origin of the measured spectra lies in the trigonal warping of the Fermi surface, implying the valley-polarization of the current jets. In addition, the demonstrated sensitivity of the collimation spectra to the trigonal warping of the Fermi surface makes such experiments a unique probe to the Fermi surface shape. %Finally, $T$-dependent measurements indicate that the jet separation remains visible up to 40~K. 
The realization of a valley-polarized current source with controllable polarization is promising for the future of valleytronic devices operating in the classical \cite{rycerz2007} and quantum \cite{tong2022} regimes. 
%\section{Methods}
%The samples were prepared following the dry transfer technique described in Refs.~\cite{zomer2014,purdie2018}, and the contacts were defined using standard e-beam lithography, evaporation, and reactive ion etching techniques, as described in Refs.~\cite{wang2013, ingla2023}.

%The electronic measurements were performed in a He cryostat at a temperature of 1.8~K unless stated otherwise and using a superconducting magnet. The electronic measurements were performed using a standard lock-in technique with a frequency of approximately 14~Hz. 

%The semi-classical simulations are described in detail in the supplementary section~6 and the codes and simulated data are available in Ref.~\cite{zenodo}.
\section*{Acknowledgements}
 We thank A.~Akhmerov, K.~Vilkelis and M.~Bozkurt for insightful discussions. This project received funding from the European Union Horizon 2020 research and innovation program under grant agreement no. 863098 (SPRING) and Marie Sklodowska–Curie individual fellowship No. 101027187-PCSV awarded to JIA. ALRM work was supported by VIDI grant 016.Vidi.189.180. KW and TT acknowledge support from JSPS KAKENHI (Grant Numbers 21H05233 and 23H02052) and World Premier International Research Center Initiative (WPI), MEXT, Japan.
%\section{Author contributions}
%JIA, ARLM and HSJvdZ conceived the project. JIA performed the measurements with help from TSG. ARLM performed the simulations with help from JIA. KW and TT synthesized the hexagonal boron nitride crystals. JIA wrote the manuscript with inputs from all authors. HSJvdZ supervised the project.
\vspace{10pt}
\section*{Data availability} 
All the data and code associated with the analysis and theoretical simulations are available free of charge from Ref.~\cite{zenodo}.
\bibliography{bibliography}

\begin{thebibliography}{43}%
\makeatletter
\providecommand \@ifxundefined [1]{%
 \@ifx{#1\undefined}
}%
\providecommand \@ifnum [1]{%
 \ifnum #1\expandafter \@firstoftwo
 \else \expandafter \@secondoftwo
 \fi
}%
\providecommand \@ifx [1]{%
 \ifx #1\expandafter \@firstoftwo
 \else \expandafter \@secondoftwo
 \fi
}%
\providecommand \natexlab [1]{#1}%
\providecommand \enquote  [1]{``#1''}%
\providecommand \bibnamefont  [1]{#1}%
\providecommand \bibfnamefont [1]{#1}%
\providecommand \citenamefont [1]{#1}%
\providecommand \href@noop [0]{\@secondoftwo}%
\providecommand \href [0]{\begingroup \@sanitize@url \@href}%
\providecommand \@href[1]{\@@startlink{#1}\@@href}%
\providecommand \@@href[1]{\endgroup#1\@@endlink}%
\providecommand \@sanitize@url [0]{\catcode `\\12\catcode `\$12\catcode
  `\&12\catcode `\#12\catcode `\^12\catcode `\_12\catcode `\%12\relax}%
\providecommand \@@startlink[1]{}%
\providecommand \@@endlink[0]{}%
\providecommand \url  [0]{\begingroup\@sanitize@url \@url }%
\providecommand \@url [1]{\endgroup\@href {#1}{\urlprefix }}%
\providecommand \urlprefix  [0]{URL }%
\providecommand \Eprint [0]{\href }%
\providecommand \doibase [0]{https://doi.org/}%
\providecommand \selectlanguage [0]{\@gobble}%
\providecommand \bibinfo  [0]{\@secondoftwo}%
\providecommand \bibfield  [0]{\@secondoftwo}%
\providecommand \translation [1]{[#1]}%
\providecommand \BibitemOpen [0]{}%
\providecommand \bibitemStop [0]{}%
\providecommand \bibitemNoStop [0]{.\EOS\space}%
\providecommand \EOS [0]{\spacefactor3000\relax}%
\providecommand \BibitemShut  [1]{\csname bibitem#1\endcsname}%
\let\auto@bib@innerbib\@empty
%</preamble>
\bibitem [{\citenamefont {Schaibley}\ \emph {et~al.}(2016)\citenamefont
  {Schaibley}, \citenamefont {Yu}, \citenamefont {Clark}, \citenamefont
  {Rivera}, \citenamefont {Ross}, \citenamefont {Seyler}, \citenamefont {Yao},\
  and\ \citenamefont {Xu}}]{schaibley2016}%
  \BibitemOpen
  \bibfield  {author} {\bibinfo {author} {\bibfnamefont {J.~R.}\ \bibnamefont
  {Schaibley}}, \bibinfo {author} {\bibfnamefont {H.}~\bibnamefont {Yu}},
  \bibinfo {author} {\bibfnamefont {G.}~\bibnamefont {Clark}}, \bibinfo
  {author} {\bibfnamefont {P.}~\bibnamefont {Rivera}}, \bibinfo {author}
  {\bibfnamefont {J.~S.}\ \bibnamefont {Ross}}, \bibinfo {author}
  {\bibfnamefont {K.~L.}\ \bibnamefont {Seyler}}, \bibinfo {author}
  {\bibfnamefont {W.}~\bibnamefont {Yao}},\ and\ \bibinfo {author}
  {\bibfnamefont {X.}~\bibnamefont {Xu}},\ }\bibfield  {title} {\bibinfo
  {title} {Valleytronics in 2d materials},\ }\href@noop {} {\bibfield
  {journal} {\bibinfo  {journal} {Nature Reviews Materials}\ }\textbf {\bibinfo
  {volume} {1}},\ \bibinfo {pages} {1} (\bibinfo {year} {2016})}\BibitemShut
  {NoStop}%
\bibitem [{\citenamefont {Yu}\ \emph {et~al.}(2020)\citenamefont {Yu},
  \citenamefont {Guan}, \citenamefont {Sheng}, \citenamefont {Gao},\ and\
  \citenamefont {Yang}}]{yu2020}%
  \BibitemOpen
  \bibfield  {author} {\bibinfo {author} {\bibfnamefont {Z.-M.}\ \bibnamefont
  {Yu}}, \bibinfo {author} {\bibfnamefont {S.}~\bibnamefont {Guan}}, \bibinfo
  {author} {\bibfnamefont {X.-L.}\ \bibnamefont {Sheng}}, \bibinfo {author}
  {\bibfnamefont {W.}~\bibnamefont {Gao}},\ and\ \bibinfo {author}
  {\bibfnamefont {S.~A.}\ \bibnamefont {Yang}},\ }\bibfield  {title} {\bibinfo
  {title} {Valley-layer coupling: a new design principle for valleytronics},\
  }\href@noop {} {\bibfield  {journal} {\bibinfo  {journal} {Physical Review
  Letters}\ }\textbf {\bibinfo {volume} {124}},\ \bibinfo {pages} {037701}
  (\bibinfo {year} {2020})}\BibitemShut {NoStop}%
\bibitem [{\citenamefont {Martin}\ \emph {et~al.}(2008)\citenamefont {Martin},
  \citenamefont {Blanter},\ and\ \citenamefont {Morpurgo}}]{martin2008}%
  \BibitemOpen
  \bibfield  {author} {\bibinfo {author} {\bibfnamefont {I.}~\bibnamefont
  {Martin}}, \bibinfo {author} {\bibfnamefont {Y.~M.}\ \bibnamefont
  {Blanter}},\ and\ \bibinfo {author} {\bibfnamefont {A.}~\bibnamefont
  {Morpurgo}},\ }\bibfield  {title} {\bibinfo {title} {Topological confinement
  in bilayer graphene},\ }\href@noop {} {\bibfield  {journal} {\bibinfo
  {journal} {Physical Review Letters}\ }\textbf {\bibinfo {volume} {100}},\
  \bibinfo {pages} {036804} (\bibinfo {year} {2008})}\BibitemShut {NoStop}%
\bibitem [{\citenamefont {San-Jose}\ \emph {et~al.}(2009)\citenamefont
  {San-Jose}, \citenamefont {Prada}, \citenamefont {McCann},\ and\
  \citenamefont {Schomerus}}]{san2009}%
  \BibitemOpen
  \bibfield  {author} {\bibinfo {author} {\bibfnamefont {P.}~\bibnamefont
  {San-Jose}}, \bibinfo {author} {\bibfnamefont {E.}~\bibnamefont {Prada}},
  \bibinfo {author} {\bibfnamefont {E.}~\bibnamefont {McCann}},\ and\ \bibinfo
  {author} {\bibfnamefont {H.}~\bibnamefont {Schomerus}},\ }\bibfield  {title}
  {\bibinfo {title} {Pseudospin valve in bilayer graphene: towards
  graphene-based pseudospintronics},\ }\href@noop {} {\bibfield  {journal}
  {\bibinfo  {journal} {Physical Review Letters}\ }\textbf {\bibinfo {volume}
  {102}},\ \bibinfo {pages} {247204} (\bibinfo {year} {2009})}\BibitemShut
  {NoStop}%
\bibitem [{\citenamefont {Li}\ \emph {et~al.}(2016)\citenamefont {Li},
  \citenamefont {Wang}, \citenamefont {McFaul}, \citenamefont {Zern},
  \citenamefont {Ren}, \citenamefont {Watanabe}, \citenamefont {Taniguchi},
  \citenamefont {Qiao},\ and\ \citenamefont {Zhu}}]{li2016}%
  \BibitemOpen
  \bibfield  {author} {\bibinfo {author} {\bibfnamefont {J.}~\bibnamefont
  {Li}}, \bibinfo {author} {\bibfnamefont {K.}~\bibnamefont {Wang}}, \bibinfo
  {author} {\bibfnamefont {K.~J.}\ \bibnamefont {McFaul}}, \bibinfo {author}
  {\bibfnamefont {Z.}~\bibnamefont {Zern}}, \bibinfo {author} {\bibfnamefont
  {Y.}~\bibnamefont {Ren}}, \bibinfo {author} {\bibfnamefont {K.}~\bibnamefont
  {Watanabe}}, \bibinfo {author} {\bibfnamefont {T.}~\bibnamefont {Taniguchi}},
  \bibinfo {author} {\bibfnamefont {Z.}~\bibnamefont {Qiao}},\ and\ \bibinfo
  {author} {\bibfnamefont {J.}~\bibnamefont {Zhu}},\ }\bibfield  {title}
  {\bibinfo {title} {Gate-controlled topological conducting channels in bilayer
  graphene},\ }\href@noop {} {\bibfield  {journal} {\bibinfo  {journal} {Nature
  Nanotechnology}\ }\textbf {\bibinfo {volume} {11}},\ \bibinfo {pages} {1060}
  (\bibinfo {year} {2016})}\BibitemShut {NoStop}%
\bibitem [{\citenamefont {Li}\ \emph {et~al.}(2018)\citenamefont {Li},
  \citenamefont {Zhang}, \citenamefont {Yin}, \citenamefont {Zhang},
  \citenamefont {Watanabe}, \citenamefont {Taniguchi}, \citenamefont {Liu},\
  and\ \citenamefont {Zhu}}]{li2018}%
  \BibitemOpen
  \bibfield  {author} {\bibinfo {author} {\bibfnamefont {J.}~\bibnamefont
  {Li}}, \bibinfo {author} {\bibfnamefont {R.-X.}\ \bibnamefont {Zhang}},
  \bibinfo {author} {\bibfnamefont {Z.}~\bibnamefont {Yin}}, \bibinfo {author}
  {\bibfnamefont {J.}~\bibnamefont {Zhang}}, \bibinfo {author} {\bibfnamefont
  {K.}~\bibnamefont {Watanabe}}, \bibinfo {author} {\bibfnamefont
  {T.}~\bibnamefont {Taniguchi}}, \bibinfo {author} {\bibfnamefont
  {C.}~\bibnamefont {Liu}},\ and\ \bibinfo {author} {\bibfnamefont
  {J.}~\bibnamefont {Zhu}},\ }\bibfield  {title} {\bibinfo {title} {A valley
  valve and electron beam splitter},\ }\href@noop {} {\bibfield  {journal}
  {\bibinfo  {journal} {Science}\ }\textbf {\bibinfo {volume} {362}},\ \bibinfo
  {pages} {1149} (\bibinfo {year} {2018})}\BibitemShut {NoStop}%
\bibitem [{\citenamefont {Yin}\ \emph {et~al.}(2022)\citenamefont {Yin},
  \citenamefont {Tan}, \citenamefont {Barcons-Ruiz}, \citenamefont {Torre},
  \citenamefont {Watanabe}, \citenamefont {Taniguchi}, \citenamefont {Song},
  \citenamefont {Hone},\ and\ \citenamefont {Koppens}}]{yin2022}%
  \BibitemOpen
  \bibfield  {author} {\bibinfo {author} {\bibfnamefont {J.}~\bibnamefont
  {Yin}}, \bibinfo {author} {\bibfnamefont {C.}~\bibnamefont {Tan}}, \bibinfo
  {author} {\bibfnamefont {D.}~\bibnamefont {Barcons-Ruiz}}, \bibinfo {author}
  {\bibfnamefont {I.}~\bibnamefont {Torre}}, \bibinfo {author} {\bibfnamefont
  {K.}~\bibnamefont {Watanabe}}, \bibinfo {author} {\bibfnamefont
  {T.}~\bibnamefont {Taniguchi}}, \bibinfo {author} {\bibfnamefont {J.~C.}\
  \bibnamefont {Song}}, \bibinfo {author} {\bibfnamefont {J.}~\bibnamefont
  {Hone}},\ and\ \bibinfo {author} {\bibfnamefont {F.~H.}\ \bibnamefont
  {Koppens}},\ }\bibfield  {title} {\bibinfo {title} {Tunable and giant
  valley-selective hall effect in gapped bilayer graphene},\ }\href@noop {}
  {\bibfield  {journal} {\bibinfo  {journal} {Science}\ }\textbf {\bibinfo
  {volume} {375}},\ \bibinfo {pages} {1398} (\bibinfo {year}
  {2022})}\BibitemShut {NoStop}%
\bibitem [{\citenamefont {Castro}\ \emph {et~al.}(2007)\citenamefont {Castro},
  \citenamefont {Novoselov}, \citenamefont {Morozov}, \citenamefont {Peres},
  \citenamefont {Dos~Santos}, \citenamefont {Nilsson}, \citenamefont {Guinea},
  \citenamefont {Geim},\ and\ \citenamefont {Neto}}]{castro2007}%
  \BibitemOpen
  \bibfield  {author} {\bibinfo {author} {\bibfnamefont {E.~V.}\ \bibnamefont
  {Castro}}, \bibinfo {author} {\bibfnamefont {K.}~\bibnamefont {Novoselov}},
  \bibinfo {author} {\bibfnamefont {S.}~\bibnamefont {Morozov}}, \bibinfo
  {author} {\bibfnamefont {N.}~\bibnamefont {Peres}}, \bibinfo {author}
  {\bibfnamefont {J.~L.}\ \bibnamefont {Dos~Santos}}, \bibinfo {author}
  {\bibfnamefont {J.}~\bibnamefont {Nilsson}}, \bibinfo {author} {\bibfnamefont
  {F.}~\bibnamefont {Guinea}}, \bibinfo {author} {\bibfnamefont
  {A.}~\bibnamefont {Geim}},\ and\ \bibinfo {author} {\bibfnamefont {A.~C.}\
  \bibnamefont {Neto}},\ }\bibfield  {title} {\bibinfo {title} {Biased bilayer
  graphene: semiconductor with a gap tunable by the electric field effect},\
  }\href@noop {} {\bibfield  {journal} {\bibinfo  {journal} {Physical Review
  Letters}\ }\textbf {\bibinfo {volume} {99}},\ \bibinfo {pages} {216802}
  (\bibinfo {year} {2007})}\BibitemShut {NoStop}%
\bibitem [{\citenamefont {Oostinga}\ \emph {et~al.}(2008)\citenamefont
  {Oostinga}, \citenamefont {Heersche}, \citenamefont {Liu}, \citenamefont
  {Morpurgo},\ and\ \citenamefont {Vandersypen}}]{oostinga2008}%
  \BibitemOpen
  \bibfield  {author} {\bibinfo {author} {\bibfnamefont {J.~B.}\ \bibnamefont
  {Oostinga}}, \bibinfo {author} {\bibfnamefont {H.~B.}\ \bibnamefont
  {Heersche}}, \bibinfo {author} {\bibfnamefont {X.}~\bibnamefont {Liu}},
  \bibinfo {author} {\bibfnamefont {A.~F.}\ \bibnamefont {Morpurgo}},\ and\
  \bibinfo {author} {\bibfnamefont {L.~M.}\ \bibnamefont {Vandersypen}},\
  }\bibfield  {title} {\bibinfo {title} {Gate-induced insulating state in
  bilayer graphene devices},\ }\href@noop {} {\bibfield  {journal} {\bibinfo
  {journal} {Nature Materials}\ }\textbf {\bibinfo {volume} {7}},\ \bibinfo
  {pages} {151} (\bibinfo {year} {2008})}\BibitemShut {NoStop}%
\bibitem [{\citenamefont {Zhang}\ \emph {et~al.}(2009)\citenamefont {Zhang},
  \citenamefont {Tang}, \citenamefont {Girit}, \citenamefont {Hao},
  \citenamefont {Martin}, \citenamefont {Zettl}, \citenamefont {Crommie},
  \citenamefont {Shen},\ and\ \citenamefont {Wang}}]{zhang2009}%
  \BibitemOpen
  \bibfield  {author} {\bibinfo {author} {\bibfnamefont {Y.}~\bibnamefont
  {Zhang}}, \bibinfo {author} {\bibfnamefont {T.-T.}\ \bibnamefont {Tang}},
  \bibinfo {author} {\bibfnamefont {C.}~\bibnamefont {Girit}}, \bibinfo
  {author} {\bibfnamefont {Z.}~\bibnamefont {Hao}}, \bibinfo {author}
  {\bibfnamefont {M.~C.}\ \bibnamefont {Martin}}, \bibinfo {author}
  {\bibfnamefont {A.}~\bibnamefont {Zettl}}, \bibinfo {author} {\bibfnamefont
  {M.~F.}\ \bibnamefont {Crommie}}, \bibinfo {author} {\bibfnamefont {Y.~R.}\
  \bibnamefont {Shen}},\ and\ \bibinfo {author} {\bibfnamefont
  {F.}~\bibnamefont {Wang}},\ }\bibfield  {title} {\bibinfo {title} {Direct
  observation of a widely tunable bandgap in bilayer graphene},\ }\href@noop {}
  {\bibfield  {journal} {\bibinfo  {journal} {Nature}\ }\textbf {\bibinfo
  {volume} {459}},\ \bibinfo {pages} {820} (\bibinfo {year}
  {2009})}\BibitemShut {NoStop}%
\bibitem [{\citenamefont {Icking}\ \emph {et~al.}(2022)\citenamefont {Icking},
  \citenamefont {Banszerus}, \citenamefont {W{\"o}rtche}, \citenamefont
  {Volmer}, \citenamefont {Schmidt}, \citenamefont {Steiner}, \citenamefont
  {Engels}, \citenamefont {Hesselmann}, \citenamefont {Goldsche}, \citenamefont
  {Watanabe} \emph {et~al.}}]{icking2022}%
  \BibitemOpen
  \bibfield  {author} {\bibinfo {author} {\bibfnamefont {E.}~\bibnamefont
  {Icking}}, \bibinfo {author} {\bibfnamefont {L.}~\bibnamefont {Banszerus}},
  \bibinfo {author} {\bibfnamefont {F.}~\bibnamefont {W{\"o}rtche}}, \bibinfo
  {author} {\bibfnamefont {F.}~\bibnamefont {Volmer}}, \bibinfo {author}
  {\bibfnamefont {P.}~\bibnamefont {Schmidt}}, \bibinfo {author} {\bibfnamefont
  {C.}~\bibnamefont {Steiner}}, \bibinfo {author} {\bibfnamefont
  {S.}~\bibnamefont {Engels}}, \bibinfo {author} {\bibfnamefont
  {J.}~\bibnamefont {Hesselmann}}, \bibinfo {author} {\bibfnamefont
  {M.}~\bibnamefont {Goldsche}}, \bibinfo {author} {\bibfnamefont
  {K.}~\bibnamefont {Watanabe}}, \emph {et~al.},\ }\bibfield  {title} {\bibinfo
  {title} {Transport spectroscopy of ultraclean tunable band gaps in bilayer
  graphene},\ }\href@noop {} {\bibfield  {journal} {\bibinfo  {journal}
  {Advanced Electronic Materials}\ }\textbf {\bibinfo {volume} {8}},\ \bibinfo
  {pages} {2200510} (\bibinfo {year} {2022})}\BibitemShut {NoStop}%
\bibitem [{\citenamefont {McCann}\ and\ \citenamefont
  {Koshino}(2013)}]{mccann2013}%
  \BibitemOpen
  \bibfield  {author} {\bibinfo {author} {\bibfnamefont {E.}~\bibnamefont
  {McCann}}\ and\ \bibinfo {author} {\bibfnamefont {M.}~\bibnamefont
  {Koshino}},\ }\bibfield  {title} {\bibinfo {title} {The electronic properties
  of bilayer graphene},\ }\href@noop {} {\bibfield  {journal} {\bibinfo
  {journal} {Reports on Progress in Physics}\ }\textbf {\bibinfo {volume}
  {76}},\ \bibinfo {pages} {056503} (\bibinfo {year} {2013})}\BibitemShut
  {NoStop}%
\bibitem [{\citenamefont {Garcia-Pomar}\ \emph {et~al.}(2008)\citenamefont
  {Garcia-Pomar}, \citenamefont {Cortijo},\ and\ \citenamefont
  {Nieto-Vesperinas}}]{garcia2008}%
  \BibitemOpen
  \bibfield  {author} {\bibinfo {author} {\bibfnamefont {J.}~\bibnamefont
  {Garcia-Pomar}}, \bibinfo {author} {\bibfnamefont {A.}~\bibnamefont
  {Cortijo}},\ and\ \bibinfo {author} {\bibfnamefont {M.}~\bibnamefont
  {Nieto-Vesperinas}},\ }\bibfield  {title} {\bibinfo {title} {Fully
  valley-polarized electron beams in graphene},\ }\href@noop {} {\bibfield
  {journal} {\bibinfo  {journal} {Physical Review Letters}\ }\textbf {\bibinfo
  {volume} {100}},\ \bibinfo {pages} {236801} (\bibinfo {year}
  {2008})}\BibitemShut {NoStop}%
\bibitem [{\citenamefont {Wang}\ and\ \citenamefont {Liu}(2010)}]{wang2010}%
  \BibitemOpen
  \bibfield  {author} {\bibinfo {author} {\bibfnamefont {Z.}~\bibnamefont
  {Wang}}\ and\ \bibinfo {author} {\bibfnamefont {F.}~\bibnamefont {Liu}},\
  }\bibfield  {title} {\bibinfo {title} {Manipulation of electron beam
  propagation by hetero-dimensional graphene junctions},\ }\href@noop {}
  {\bibfield  {journal} {\bibinfo  {journal} {ACS Nano}\ }\textbf {\bibinfo
  {volume} {4}},\ \bibinfo {pages} {2459} (\bibinfo {year} {2010})}\BibitemShut
  {NoStop}%
\bibitem [{\citenamefont {Stegmann}\ and\ \citenamefont
  {Szpak}(2018)}]{stegmann2018}%
  \BibitemOpen
  \bibfield  {author} {\bibinfo {author} {\bibfnamefont {T.}~\bibnamefont
  {Stegmann}}\ and\ \bibinfo {author} {\bibfnamefont {N.}~\bibnamefont
  {Szpak}},\ }\bibfield  {title} {\bibinfo {title} {Current splitting and
  valley polarization in elastically deformed graphene},\ }\href@noop {}
  {\bibfield  {journal} {\bibinfo  {journal} {2D Materials}\ }\textbf {\bibinfo
  {volume} {6}},\ \bibinfo {pages} {015024} (\bibinfo {year}
  {2018})}\BibitemShut {NoStop}%
\bibitem [{\citenamefont {Manesco}\ and\ \citenamefont
  {Pulkin}(2023)}]{manesco2023}%
  \BibitemOpen
  \bibfield  {author} {\bibinfo {author} {\bibfnamefont {A.~L.}\ \bibnamefont
  {Manesco}}\ and\ \bibinfo {author} {\bibfnamefont {A.}~\bibnamefont
  {Pulkin}},\ }\bibfield  {title} {\bibinfo {title} {Spatial separation of spin
  currents in transition metal dichalcogenides},\ }\href@noop {} {\bibfield
  {journal} {\bibinfo  {journal} {SciPost Physics Core}\ }\textbf {\bibinfo
  {volume} {6}},\ \bibinfo {pages} {036} (\bibinfo {year} {2023})}\BibitemShut
  {NoStop}%
\bibitem [{\citenamefont {Gold}\ \emph {et~al.}(2021)\citenamefont {Gold},
  \citenamefont {Knothe}, \citenamefont {Kurzmann}, \citenamefont
  {Garcia-Ruiz}, \citenamefont {Watanabe}, \citenamefont {Taniguchi},
  \citenamefont {Fal’ko}, \citenamefont {Ensslin},\ and\ \citenamefont
  {Ihn}}]{gold2021}%
  \BibitemOpen
  \bibfield  {author} {\bibinfo {author} {\bibfnamefont {C.}~\bibnamefont
  {Gold}}, \bibinfo {author} {\bibfnamefont {A.}~\bibnamefont {Knothe}},
  \bibinfo {author} {\bibfnamefont {A.}~\bibnamefont {Kurzmann}}, \bibinfo
  {author} {\bibfnamefont {A.}~\bibnamefont {Garcia-Ruiz}}, \bibinfo {author}
  {\bibfnamefont {K.}~\bibnamefont {Watanabe}}, \bibinfo {author}
  {\bibfnamefont {T.}~\bibnamefont {Taniguchi}}, \bibinfo {author}
  {\bibfnamefont {V.}~\bibnamefont {Fal’ko}}, \bibinfo {author}
  {\bibfnamefont {K.}~\bibnamefont {Ensslin}},\ and\ \bibinfo {author}
  {\bibfnamefont {T.}~\bibnamefont {Ihn}},\ }\bibfield  {title} {\bibinfo
  {title} {Coherent jetting from a gate-defined channel in bilayer graphene},\
  }\href@noop {} {\bibfield  {journal} {\bibinfo  {journal} {Physical Review
  Letters}\ }\textbf {\bibinfo {volume} {127}},\ \bibinfo {pages} {046801}
  (\bibinfo {year} {2021})}\BibitemShut {NoStop}%
\bibitem [{\citenamefont {Overweg}\ \emph
  {et~al.}(2018{\natexlab{a}})\citenamefont {Overweg}, \citenamefont
  {Eggimann}, \citenamefont {Chen}, \citenamefont {Slizovskiy}, \citenamefont
  {Eich}, \citenamefont {Pisoni}, \citenamefont {Lee}, \citenamefont
  {Rickhaus}, \citenamefont {Watanabe}, \citenamefont {Taniguchi} \emph
  {et~al.}}]{1overweg2018}%
  \BibitemOpen
  \bibfield  {author} {\bibinfo {author} {\bibfnamefont {H.}~\bibnamefont
  {Overweg}}, \bibinfo {author} {\bibfnamefont {H.}~\bibnamefont {Eggimann}},
  \bibinfo {author} {\bibfnamefont {X.}~\bibnamefont {Chen}}, \bibinfo {author}
  {\bibfnamefont {S.}~\bibnamefont {Slizovskiy}}, \bibinfo {author}
  {\bibfnamefont {M.}~\bibnamefont {Eich}}, \bibinfo {author} {\bibfnamefont
  {R.}~\bibnamefont {Pisoni}}, \bibinfo {author} {\bibfnamefont
  {Y.}~\bibnamefont {Lee}}, \bibinfo {author} {\bibfnamefont {P.}~\bibnamefont
  {Rickhaus}}, \bibinfo {author} {\bibfnamefont {K.}~\bibnamefont {Watanabe}},
  \bibinfo {author} {\bibfnamefont {T.}~\bibnamefont {Taniguchi}}, \emph
  {et~al.},\ }\bibfield  {title} {\bibinfo {title} {Electrostatically induced
  quantum point contacts in bilayer graphene},\ }\href@noop {} {\bibfield
  {journal} {\bibinfo  {journal} {Nano Letters}\ }\textbf {\bibinfo {volume}
  {18}},\ \bibinfo {pages} {553} (\bibinfo {year}
  {2018}{\natexlab{a}})}\BibitemShut {NoStop}%
\bibitem [{\citenamefont {Kraft}\ \emph
  {et~al.}(2018{\natexlab{a}})\citenamefont {Kraft}, \citenamefont {Mohrmann},
  \citenamefont {Du}, \citenamefont {Selvasundaram}, \citenamefont {Irfan},
  \citenamefont {Kanilmaz}, \citenamefont {Wu}, \citenamefont {Beckmann},
  \citenamefont {Von~L{\"o}hneysen}, \citenamefont {Krupke} \emph
  {et~al.}}]{1kraft2018}%
  \BibitemOpen
  \bibfield  {author} {\bibinfo {author} {\bibfnamefont {R.}~\bibnamefont
  {Kraft}}, \bibinfo {author} {\bibfnamefont {J.}~\bibnamefont {Mohrmann}},
  \bibinfo {author} {\bibfnamefont {R.}~\bibnamefont {Du}}, \bibinfo {author}
  {\bibfnamefont {P.~B.}\ \bibnamefont {Selvasundaram}}, \bibinfo {author}
  {\bibfnamefont {M.}~\bibnamefont {Irfan}}, \bibinfo {author} {\bibfnamefont
  {U.~N.}\ \bibnamefont {Kanilmaz}}, \bibinfo {author} {\bibfnamefont
  {F.}~\bibnamefont {Wu}}, \bibinfo {author} {\bibfnamefont {D.}~\bibnamefont
  {Beckmann}}, \bibinfo {author} {\bibfnamefont {H.}~\bibnamefont
  {Von~L{\"o}hneysen}}, \bibinfo {author} {\bibfnamefont {R.}~\bibnamefont
  {Krupke}}, \emph {et~al.},\ }\bibfield  {title} {\bibinfo {title} {Tailoring
  supercurrent confinement in graphene bilayer weak links},\ }\href@noop {}
  {\bibfield  {journal} {\bibinfo  {journal} {Nature Communications}\ }\textbf
  {\bibinfo {volume} {9}},\ \bibinfo {pages} {1722} (\bibinfo {year}
  {2018}{\natexlab{a}})}\BibitemShut {NoStop}%
\bibitem [{\citenamefont {Overweg}\ \emph
  {et~al.}(2018{\natexlab{b}})\citenamefont {Overweg}, \citenamefont {Knothe},
  \citenamefont {Fabian}, \citenamefont {Linhart}, \citenamefont {Rickhaus},
  \citenamefont {Wernli}, \citenamefont {Watanabe}, \citenamefont {Taniguchi},
  \citenamefont {S{\'a}nchez}, \citenamefont {Burgd{\"o}rfer} \emph
  {et~al.}}]{overweg2018}%
  \BibitemOpen
  \bibfield  {author} {\bibinfo {author} {\bibfnamefont {H.}~\bibnamefont
  {Overweg}}, \bibinfo {author} {\bibfnamefont {A.}~\bibnamefont {Knothe}},
  \bibinfo {author} {\bibfnamefont {T.}~\bibnamefont {Fabian}}, \bibinfo
  {author} {\bibfnamefont {L.}~\bibnamefont {Linhart}}, \bibinfo {author}
  {\bibfnamefont {P.}~\bibnamefont {Rickhaus}}, \bibinfo {author}
  {\bibfnamefont {L.}~\bibnamefont {Wernli}}, \bibinfo {author} {\bibfnamefont
  {K.}~\bibnamefont {Watanabe}}, \bibinfo {author} {\bibfnamefont
  {T.}~\bibnamefont {Taniguchi}}, \bibinfo {author} {\bibfnamefont
  {D.}~\bibnamefont {S{\'a}nchez}}, \bibinfo {author} {\bibfnamefont
  {J.}~\bibnamefont {Burgd{\"o}rfer}}, \emph {et~al.},\ }\bibfield  {title}
  {\bibinfo {title} {Topologically nontrivial valley states in bilayer graphene
  quantum point contacts},\ }\href@noop {} {\bibfield  {journal} {\bibinfo
  {journal} {Physical Review Letters}\ }\textbf {\bibinfo {volume} {121}},\
  \bibinfo {pages} {257702} (\bibinfo {year} {2018}{\natexlab{b}})}\BibitemShut
  {NoStop}%
\bibitem [{\citenamefont {Kraft}\ \emph
  {et~al.}(2018{\natexlab{b}})\citenamefont {Kraft}, \citenamefont {Krainov},
  \citenamefont {Gall}, \citenamefont {Dmitriev}, \citenamefont {Krupke},
  \citenamefont {Gornyi},\ and\ \citenamefont {Danneau}}]{kraft2018}%
  \BibitemOpen
  \bibfield  {author} {\bibinfo {author} {\bibfnamefont {R.}~\bibnamefont
  {Kraft}}, \bibinfo {author} {\bibfnamefont {I.}~\bibnamefont {Krainov}},
  \bibinfo {author} {\bibfnamefont {V.}~\bibnamefont {Gall}}, \bibinfo {author}
  {\bibfnamefont {A.}~\bibnamefont {Dmitriev}}, \bibinfo {author}
  {\bibfnamefont {R.}~\bibnamefont {Krupke}}, \bibinfo {author} {\bibfnamefont
  {I.}~\bibnamefont {Gornyi}},\ and\ \bibinfo {author} {\bibfnamefont
  {R.}~\bibnamefont {Danneau}},\ }\bibfield  {title} {\bibinfo {title} {Valley
  subband splitting in bilayer graphene quantum point contacts},\ }\href@noop
  {} {\bibfield  {journal} {\bibinfo  {journal} {Physical Review Letters}\
  }\textbf {\bibinfo {volume} {121}},\ \bibinfo {pages} {257703} (\bibinfo
  {year} {2018}{\natexlab{b}})}\BibitemShut {NoStop}%
\bibitem [{\citenamefont {Knothe}\ and\ \citenamefont
  {Fal'ko}(2018)}]{knothe2018}%
  \BibitemOpen
  \bibfield  {author} {\bibinfo {author} {\bibfnamefont {A.}~\bibnamefont
  {Knothe}}\ and\ \bibinfo {author} {\bibfnamefont {V.}~\bibnamefont
  {Fal'ko}},\ }\bibfield  {title} {\bibinfo {title} {Influence of minivalleys
  and berry curvature on electrostatically induced quantum wires in gapped
  bilayer graphene},\ }\href@noop {} {\bibfield  {journal} {\bibinfo  {journal}
  {Physical Review B}\ }\textbf {\bibinfo {volume} {98}},\ \bibinfo {pages}
  {155435} (\bibinfo {year} {2018})}\BibitemShut {NoStop}%
\bibitem [{\citenamefont {Lane}\ \emph {et~al.}(2019)\citenamefont {Lane},
  \citenamefont {Knothe},\ and\ \citenamefont {Fal'ko}}]{lane2019}%
  \BibitemOpen
  \bibfield  {author} {\bibinfo {author} {\bibfnamefont {T.~L.}\ \bibnamefont
  {Lane}}, \bibinfo {author} {\bibfnamefont {A.}~\bibnamefont {Knothe}},\ and\
  \bibinfo {author} {\bibfnamefont {V.~I.}\ \bibnamefont {Fal'ko}},\ }\bibfield
   {title} {\bibinfo {title} {Semimetallic features in quantum transport
  through a gate-defined point contact in bilayer graphene},\ }\href@noop {}
  {\bibfield  {journal} {\bibinfo  {journal} {Physical Review B}\ }\textbf
  {\bibinfo {volume} {100}},\ \bibinfo {pages} {115427} (\bibinfo {year}
  {2019})}\BibitemShut {NoStop}%
\bibitem [{\citenamefont {Banszerus}\ \emph {et~al.}(2020)\citenamefont
  {Banszerus}, \citenamefont {Frohn}, \citenamefont {Fabian}, \citenamefont
  {Somanchi}, \citenamefont {Epping}, \citenamefont {M{\"u}ller}, \citenamefont
  {Neumaier}, \citenamefont {Watanabe}, \citenamefont {Taniguchi},
  \citenamefont {Libisch} \emph {et~al.}}]{banszerus2020}%
  \BibitemOpen
  \bibfield  {author} {\bibinfo {author} {\bibfnamefont {L.}~\bibnamefont
  {Banszerus}}, \bibinfo {author} {\bibfnamefont {B.}~\bibnamefont {Frohn}},
  \bibinfo {author} {\bibfnamefont {T.}~\bibnamefont {Fabian}}, \bibinfo
  {author} {\bibfnamefont {S.}~\bibnamefont {Somanchi}}, \bibinfo {author}
  {\bibfnamefont {A.}~\bibnamefont {Epping}}, \bibinfo {author} {\bibfnamefont
  {M.}~\bibnamefont {M{\"u}ller}}, \bibinfo {author} {\bibfnamefont
  {D.}~\bibnamefont {Neumaier}}, \bibinfo {author} {\bibfnamefont
  {K.}~\bibnamefont {Watanabe}}, \bibinfo {author} {\bibfnamefont
  {T.}~\bibnamefont {Taniguchi}}, \bibinfo {author} {\bibfnamefont
  {F.}~\bibnamefont {Libisch}}, \emph {et~al.},\ }\bibfield  {title} {\bibinfo
  {title} {Observation of the spin-orbit gap in bilayer graphene by
  one-dimensional ballistic transport},\ }\href@noop {} {\bibfield  {journal}
  {\bibinfo  {journal} {Physical Review Letters}\ }\textbf {\bibinfo {volume}
  {124}},\ \bibinfo {pages} {177701} (\bibinfo {year} {2020})}\BibitemShut
  {NoStop}%
\bibitem [{\citenamefont {Lee}\ \emph {et~al.}(2020)\citenamefont {Lee},
  \citenamefont {Knothe}, \citenamefont {Overweg}, \citenamefont {Eich},
  \citenamefont {Gold}, \citenamefont {Kurzmann}, \citenamefont {Klasovika},
  \citenamefont {Taniguchi}, \citenamefont {Wantanabe}, \citenamefont
  {Fal’ko} \emph {et~al.}}]{lee2020}%
  \BibitemOpen
  \bibfield  {author} {\bibinfo {author} {\bibfnamefont {Y.}~\bibnamefont
  {Lee}}, \bibinfo {author} {\bibfnamefont {A.}~\bibnamefont {Knothe}},
  \bibinfo {author} {\bibfnamefont {H.}~\bibnamefont {Overweg}}, \bibinfo
  {author} {\bibfnamefont {M.}~\bibnamefont {Eich}}, \bibinfo {author}
  {\bibfnamefont {C.}~\bibnamefont {Gold}}, \bibinfo {author} {\bibfnamefont
  {A.}~\bibnamefont {Kurzmann}}, \bibinfo {author} {\bibfnamefont
  {V.}~\bibnamefont {Klasovika}}, \bibinfo {author} {\bibfnamefont
  {T.}~\bibnamefont {Taniguchi}}, \bibinfo {author} {\bibfnamefont
  {K.}~\bibnamefont {Wantanabe}}, \bibinfo {author} {\bibfnamefont
  {V.}~\bibnamefont {Fal’ko}}, \emph {et~al.},\ }\bibfield  {title} {\bibinfo
  {title} {Tunable valley splitting due to topological orbital magnetic moment
  in bilayer graphene quantum point contacts},\ }\href@noop {} {\bibfield
  {journal} {\bibinfo  {journal} {Physical Review Letters}\ }\textbf {\bibinfo
  {volume} {124}},\ \bibinfo {pages} {126802} (\bibinfo {year}
  {2020})}\BibitemShut {NoStop}%
\bibitem [{\citenamefont {Gall}\ \emph {et~al.}(2022)\citenamefont {Gall},
  \citenamefont {Kraft}, \citenamefont {Gornyi},\ and\ \citenamefont
  {Danneau}}]{gall2022}%
  \BibitemOpen
  \bibfield  {author} {\bibinfo {author} {\bibfnamefont {V.}~\bibnamefont
  {Gall}}, \bibinfo {author} {\bibfnamefont {R.}~\bibnamefont {Kraft}},
  \bibinfo {author} {\bibfnamefont {I.~V.}\ \bibnamefont {Gornyi}},\ and\
  \bibinfo {author} {\bibfnamefont {R.}~\bibnamefont {Danneau}},\ }\bibfield
  {title} {\bibinfo {title} {Spin and valley degrees of freedom in a bilayer
  graphene quantum point contact: Zeeman splitting and interaction effects},\
  }\href@noop {} {\bibfield  {journal} {\bibinfo  {journal} {Physical Review
  Research}\ }\textbf {\bibinfo {volume} {4}},\ \bibinfo {pages} {023142}
  (\bibinfo {year} {2022})}\BibitemShut {NoStop}%
\bibitem [{\citenamefont {Ingla-Ayn{\'e}s}\ \emph {et~al.}(2023)\citenamefont
  {Ingla-Ayn{\'e}s}, \citenamefont {Manesco}, \citenamefont {Ghiasi},
  \citenamefont {Volosheniuk}, \citenamefont {Watanabe}, \citenamefont
  {Taniguchi},\ and\ \citenamefont {van~der Zant}}]{ingla2023}%
  \BibitemOpen
  \bibfield  {author} {\bibinfo {author} {\bibfnamefont {J.}~\bibnamefont
  {Ingla-Ayn{\'e}s}}, \bibinfo {author} {\bibfnamefont {A.~L.}\ \bibnamefont
  {Manesco}}, \bibinfo {author} {\bibfnamefont {T.~S.}\ \bibnamefont {Ghiasi}},
  \bibinfo {author} {\bibfnamefont {S.}~\bibnamefont {Volosheniuk}}, \bibinfo
  {author} {\bibfnamefont {K.}~\bibnamefont {Watanabe}}, \bibinfo {author}
  {\bibfnamefont {T.}~\bibnamefont {Taniguchi}},\ and\ \bibinfo {author}
  {\bibfnamefont {H.~S.}\ \bibnamefont {van~der Zant}},\ }\bibfield  {title}
  {\bibinfo {title} {Specular electron focusing between gate-defined quantum
  point contacts in bilayer graphene},\ }\href@noop {} {\bibfield  {journal}
  {\bibinfo  {journal} {Nano Letters}\ }\textbf {\bibinfo {volume} {23}},\
  \bibinfo {pages} {5453} (\bibinfo {year} {2023})}\BibitemShut {NoStop}%
\bibitem [{\citenamefont {Molenkamp}\ \emph {et~al.}(1990)\citenamefont
  {Molenkamp}, \citenamefont {Staring}, \citenamefont {Beenakker},
  \citenamefont {Eppenga}, \citenamefont {Timmering}, \citenamefont
  {Williamson}, \citenamefont {Harmans},\ and\ \citenamefont
  {Foxon}}]{molenkamp1990}%
  \BibitemOpen
  \bibfield  {author} {\bibinfo {author} {\bibfnamefont {L.}~\bibnamefont
  {Molenkamp}}, \bibinfo {author} {\bibfnamefont {A.}~\bibnamefont {Staring}},
  \bibinfo {author} {\bibfnamefont {C.}~\bibnamefont {Beenakker}}, \bibinfo
  {author} {\bibfnamefont {R.}~\bibnamefont {Eppenga}}, \bibinfo {author}
  {\bibfnamefont {C.}~\bibnamefont {Timmering}}, \bibinfo {author}
  {\bibfnamefont {J.}~\bibnamefont {Williamson}}, \bibinfo {author}
  {\bibfnamefont {C.}~\bibnamefont {Harmans}},\ and\ \bibinfo {author}
  {\bibfnamefont {C.}~\bibnamefont {Foxon}},\ }\bibfield  {title} {\bibinfo
  {title} {Electron-beam collimation with a quantum point contact},\
  }\href@noop {} {\bibfield  {journal} {\bibinfo  {journal} {Physical Review
  B}\ }\textbf {\bibinfo {volume} {41}},\ \bibinfo {pages} {1274} (\bibinfo
  {year} {1990})}\BibitemShut {NoStop}%
\bibitem [{\citenamefont {Yacoby}\ \emph {et~al.}(1991)\citenamefont {Yacoby},
  \citenamefont {Sivan}, \citenamefont {Umbach},\ and\ \citenamefont
  {Hong}}]{yacoby1991}%
  \BibitemOpen
  \bibfield  {author} {\bibinfo {author} {\bibfnamefont {A.}~\bibnamefont
  {Yacoby}}, \bibinfo {author} {\bibfnamefont {U.}~\bibnamefont {Sivan}},
  \bibinfo {author} {\bibfnamefont {C.}~\bibnamefont {Umbach}},\ and\ \bibinfo
  {author} {\bibfnamefont {J.}~\bibnamefont {Hong}},\ }\bibfield  {title}
  {\bibinfo {title} {Interference and dephasing by electron-electron
  interaction on length scales shorter than the elastic mean free path},\
  }\href@noop {} {\bibfield  {journal} {\bibinfo  {journal} {Physical Review
  Letters}\ }\textbf {\bibinfo {volume} {66}},\ \bibinfo {pages} {1938}
  (\bibinfo {year} {1991})}\BibitemShut {NoStop}%
\bibitem [{\citenamefont {Shepard}\ \emph {et~al.}(1992)\citenamefont
  {Shepard}, \citenamefont {Roukes},\ and\ \citenamefont {Van~der
  Gaag}}]{shepard1992}%
  \BibitemOpen
  \bibfield  {author} {\bibinfo {author} {\bibfnamefont {K.}~\bibnamefont
  {Shepard}}, \bibinfo {author} {\bibfnamefont {M.}~\bibnamefont {Roukes}},\
  and\ \bibinfo {author} {\bibfnamefont {B.}~\bibnamefont {Van~der Gaag}},\
  }\bibfield  {title} {\bibinfo {title} {Direct measurement of the transmission
  matrix of a mesoscopic conductor},\ }\href@noop {} {\bibfield  {journal}
  {\bibinfo  {journal} {Physical Review Letters}\ }\textbf {\bibinfo {volume}
  {68}},\ \bibinfo {pages} {2660} (\bibinfo {year} {1992})}\BibitemShut
  {NoStop}%
\bibitem [{\citenamefont {Barnard}\ \emph {et~al.}(2017)\citenamefont
  {Barnard}, \citenamefont {Hughes}, \citenamefont {Sharpe}, \citenamefont
  {Watanabe}, \citenamefont {Taniguchi},\ and\ \citenamefont
  {Goldhaber-Gordon}}]{barnard2017}%
  \BibitemOpen
  \bibfield  {author} {\bibinfo {author} {\bibfnamefont {A.~W.}\ \bibnamefont
  {Barnard}}, \bibinfo {author} {\bibfnamefont {A.}~\bibnamefont {Hughes}},
  \bibinfo {author} {\bibfnamefont {A.~L.}\ \bibnamefont {Sharpe}}, \bibinfo
  {author} {\bibfnamefont {K.}~\bibnamefont {Watanabe}}, \bibinfo {author}
  {\bibfnamefont {T.}~\bibnamefont {Taniguchi}},\ and\ \bibinfo {author}
  {\bibfnamefont {D.}~\bibnamefont {Goldhaber-Gordon}},\ }\bibfield  {title}
  {\bibinfo {title} {Absorptive pinhole collimators for ballistic dirac
  fermions in graphene},\ }\href@noop {} {\bibfield  {journal} {\bibinfo
  {journal} {Nature Communications}\ }\textbf {\bibinfo {volume} {8}},\
  \bibinfo {pages} {15418} (\bibinfo {year} {2017})}\BibitemShut {NoStop}%
\bibitem [{\citenamefont {Topinka}\ \emph {et~al.}(2000)\citenamefont
  {Topinka}, \citenamefont {LeRoy}, \citenamefont {Shaw}, \citenamefont
  {Heller}, \citenamefont {Westervelt}, \citenamefont {Maranowski},\ and\
  \citenamefont {Gossard}}]{topinka2000}%
  \BibitemOpen
  \bibfield  {author} {\bibinfo {author} {\bibfnamefont {M.}~\bibnamefont
  {Topinka}}, \bibinfo {author} {\bibfnamefont {B.~J.}\ \bibnamefont {LeRoy}},
  \bibinfo {author} {\bibfnamefont {S.}~\bibnamefont {Shaw}}, \bibinfo {author}
  {\bibfnamefont {E.}~\bibnamefont {Heller}}, \bibinfo {author} {\bibfnamefont
  {R.}~\bibnamefont {Westervelt}}, \bibinfo {author} {\bibfnamefont
  {K.}~\bibnamefont {Maranowski}},\ and\ \bibinfo {author} {\bibfnamefont
  {A.}~\bibnamefont {Gossard}},\ }\bibfield  {title} {\bibinfo {title} {Imaging
  coherent electron flow from a quantum point contact},\ }\href@noop {}
  {\bibfield  {journal} {\bibinfo  {journal} {Science}\ }\textbf {\bibinfo
  {volume} {289}},\ \bibinfo {pages} {2323} (\bibinfo {year}
  {2000})}\BibitemShut {NoStop}%
\bibitem [{\citenamefont {Kuzmenko}\ \emph {et~al.}(2009)\citenamefont
  {Kuzmenko}, \citenamefont {Crassee}, \citenamefont {Van Der~Marel},
  \citenamefont {Blake},\ and\ \citenamefont {Novoselov}}]{kuzmenko2009}%
  \BibitemOpen
  \bibfield  {author} {\bibinfo {author} {\bibfnamefont {A.}~\bibnamefont
  {Kuzmenko}}, \bibinfo {author} {\bibfnamefont {I.}~\bibnamefont {Crassee}},
  \bibinfo {author} {\bibfnamefont {D.}~\bibnamefont {Van Der~Marel}}, \bibinfo
  {author} {\bibfnamefont {P.}~\bibnamefont {Blake}},\ and\ \bibinfo {author}
  {\bibfnamefont {K.}~\bibnamefont {Novoselov}},\ }\bibfield  {title} {\bibinfo
  {title} {Determination of the gate-tunable band gap and tight-binding
  parameters in bilayer graphene using infrared spectroscopy},\ }\href@noop {}
  {\bibfield  {journal} {\bibinfo  {journal} {Physical Review B}\ }\textbf
  {\bibinfo {volume} {80}},\ \bibinfo {pages} {165406} (\bibinfo {year}
  {2009})}\BibitemShut {NoStop}%
\bibitem [{\citenamefont {Ingla-Ayn\'es}\ \emph {et~al.}(2023)\citenamefont
  {Ingla-Ayn\'es}, \citenamefont {Manesco}, \citenamefont {Ghiasi},
  \citenamefont {Kenji}, \citenamefont {Takashi},\ and\ \citenamefont {van~der
  Zant}}]{zenodo}%
  \BibitemOpen
  \bibfield  {author} {\bibinfo {author} {\bibfnamefont {J.}~\bibnamefont
  {Ingla-Ayn\'es}}, \bibinfo {author} {\bibfnamefont {A.}~\bibnamefont
  {Manesco}}, \bibinfo {author} {\bibfnamefont {T.~S.}\ \bibnamefont {Ghiasi}},
  \bibinfo {author} {\bibfnamefont {W.}~\bibnamefont {Kenji}}, \bibinfo
  {author} {\bibfnamefont {T.}~\bibnamefont {Takashi}},\ and\ \bibinfo {author}
  {\bibfnamefont {H.~S.~J.}\ \bibnamefont {van~der Zant}},\ }\href@noop {}
  {\bibinfo {title} {Data underlying the publication: A ballistic electron
  source with magnetically-controlled valley polarization in bilayer
  graphene}},\ \bibinfo {howpublished}
  {https://doi.org/10.4121/8bff6fc3-91d4-4bed-8f32-e36d1b2c1fbe} (\bibinfo
  {year} {2023})\BibitemShut {NoStop}%
\bibitem [{\citenamefont {B{\"u}ttiker}(1986)}]{buttiker1986}%
  \BibitemOpen
  \bibfield  {author} {\bibinfo {author} {\bibfnamefont {M.}~\bibnamefont
  {B{\"u}ttiker}},\ }\bibfield  {title} {\bibinfo {title} {Four-terminal
  phase-coherent conductance},\ }\href@noop {} {\bibfield  {journal} {\bibinfo
  {journal} {Physical Review Letters}\ }\textbf {\bibinfo {volume} {57}},\
  \bibinfo {pages} {1761} (\bibinfo {year} {1986})}\BibitemShut {NoStop}%
\bibitem [{\citenamefont {Buttiker}(1988)}]{buttiker1988}%
  \BibitemOpen
  \bibfield  {author} {\bibinfo {author} {\bibfnamefont {M.}~\bibnamefont
  {Buttiker}},\ }\bibfield  {title} {\bibinfo {title} {Symmetry of electrical
  conduction},\ }\href@noop {} {\bibfield  {journal} {\bibinfo  {journal} {IBM
  Journal of Research and Development}\ }\textbf {\bibinfo {volume} {32}},\
  \bibinfo {pages} {317} (\bibinfo {year} {1988})}\BibitemShut {NoStop}%
\bibitem [{\citenamefont {Kozikov}\ \emph {et~al.}(2012)\citenamefont
  {Kozikov}, \citenamefont {Horsell}, \citenamefont {McCann},\ and\
  \citenamefont {Fal'Ko}}]{kozikov2012}%
  \BibitemOpen
  \bibfield  {author} {\bibinfo {author} {\bibfnamefont {A.}~\bibnamefont
  {Kozikov}}, \bibinfo {author} {\bibfnamefont {D.}~\bibnamefont {Horsell}},
  \bibinfo {author} {\bibfnamefont {E.}~\bibnamefont {McCann}},\ and\ \bibinfo
  {author} {\bibfnamefont {V.}~\bibnamefont {Fal'Ko}},\ }\bibfield  {title}
  {\bibinfo {title} {Evidence for spin memory in the electron phase coherence
  in graphene},\ }\href@noop {} {\bibfield  {journal} {\bibinfo  {journal}
  {Physical Review B}\ }\textbf {\bibinfo {volume} {86}},\ \bibinfo {pages}
  {045436} (\bibinfo {year} {2012})}\BibitemShut {NoStop}%
\bibitem [{\citenamefont {Rycerz}\ \emph {et~al.}(2007)\citenamefont {Rycerz},
  \citenamefont {Tworzyd{\l}o},\ and\ \citenamefont {Beenakker}}]{rycerz2007}%
  \BibitemOpen
  \bibfield  {author} {\bibinfo {author} {\bibfnamefont {A.}~\bibnamefont
  {Rycerz}}, \bibinfo {author} {\bibfnamefont {J.}~\bibnamefont
  {Tworzyd{\l}o}},\ and\ \bibinfo {author} {\bibfnamefont {C.}~\bibnamefont
  {Beenakker}},\ }\bibfield  {title} {\bibinfo {title} {Valley filter and
  valley valve in graphene},\ }\href@noop {} {\bibfield  {journal} {\bibinfo
  {journal} {Nature Physics}\ }\textbf {\bibinfo {volume} {3}},\ \bibinfo
  {pages} {172} (\bibinfo {year} {2007})}\BibitemShut {NoStop}%
\bibitem [{\citenamefont {Tong}\ \emph {et~al.}(2022)\citenamefont {Tong},
  \citenamefont {Kurzmann}, \citenamefont {Garreis}, \citenamefont {Huang},
  \citenamefont {Jele}, \citenamefont {Eich}, \citenamefont {Ginzburg},
  \citenamefont {Mittag}, \citenamefont {Watanabe}, \citenamefont {Taniguchi}
  \emph {et~al.}}]{tong2022}%
  \BibitemOpen
  \bibfield  {author} {\bibinfo {author} {\bibfnamefont {C.}~\bibnamefont
  {Tong}}, \bibinfo {author} {\bibfnamefont {A.}~\bibnamefont {Kurzmann}},
  \bibinfo {author} {\bibfnamefont {R.}~\bibnamefont {Garreis}}, \bibinfo
  {author} {\bibfnamefont {W.~W.}\ \bibnamefont {Huang}}, \bibinfo {author}
  {\bibfnamefont {S.}~\bibnamefont {Jele}}, \bibinfo {author} {\bibfnamefont
  {M.}~\bibnamefont {Eich}}, \bibinfo {author} {\bibfnamefont {L.}~\bibnamefont
  {Ginzburg}}, \bibinfo {author} {\bibfnamefont {C.}~\bibnamefont {Mittag}},
  \bibinfo {author} {\bibfnamefont {K.}~\bibnamefont {Watanabe}}, \bibinfo
  {author} {\bibfnamefont {T.}~\bibnamefont {Taniguchi}}, \emph {et~al.},\
  }\bibfield  {title} {\bibinfo {title} {Pauli blockade of tunable two-electron
  spin and valley states in graphene quantum dots},\ }\href@noop {} {\bibfield
  {journal} {\bibinfo  {journal} {Physical Review Letters}\ }\textbf {\bibinfo
  {volume} {128}},\ \bibinfo {pages} {067702} (\bibinfo {year}
  {2022})}\BibitemShut {NoStop}%
\bibitem [{\citenamefont {Molenkamp}\ \emph {et~al.}(1992)\citenamefont
  {Molenkamp}, \citenamefont {Brugmans}, \citenamefont {Van~Houten},\ and\
  \citenamefont {Foxon}}]{molenkamp1992}%
  \BibitemOpen
  \bibfield  {author} {\bibinfo {author} {\bibfnamefont {L.}~\bibnamefont
  {Molenkamp}}, \bibinfo {author} {\bibfnamefont {M.}~\bibnamefont {Brugmans}},
  \bibinfo {author} {\bibfnamefont {H.}~\bibnamefont {Van~Houten}},\ and\
  \bibinfo {author} {\bibfnamefont {C.}~\bibnamefont {Foxon}},\ }\bibfield
  {title} {\bibinfo {title} {Electron-electron scattering probed by a
  collimated electron beam},\ }\href@noop {} {\bibfield  {journal} {\bibinfo
  {journal} {Semiconductor science and technology}\ }\textbf {\bibinfo {volume}
  {7}},\ \bibinfo {pages} {B228} (\bibinfo {year} {1992})}\BibitemShut
  {NoStop}%
\bibitem [{\citenamefont {Lee}\ \emph {et~al.}(2016)\citenamefont {Lee},
  \citenamefont {Wallbank}, \citenamefont {Gallagher}, \citenamefont
  {Watanabe}, \citenamefont {Taniguchi}, \citenamefont {Fal’ko},\ and\
  \citenamefont {Goldhaber-Gordon}}]{lee2016}%
  \BibitemOpen
  \bibfield  {author} {\bibinfo {author} {\bibfnamefont {M.}~\bibnamefont
  {Lee}}, \bibinfo {author} {\bibfnamefont {J.~R.}\ \bibnamefont {Wallbank}},
  \bibinfo {author} {\bibfnamefont {P.}~\bibnamefont {Gallagher}}, \bibinfo
  {author} {\bibfnamefont {K.}~\bibnamefont {Watanabe}}, \bibinfo {author}
  {\bibfnamefont {T.}~\bibnamefont {Taniguchi}}, \bibinfo {author}
  {\bibfnamefont {V.~I.}\ \bibnamefont {Fal’ko}},\ and\ \bibinfo {author}
  {\bibfnamefont {D.}~\bibnamefont {Goldhaber-Gordon}},\ }\bibfield  {title}
  {\bibinfo {title} {Ballistic miniband conduction in a graphene
  superlattice},\ }\href@noop {} {\bibfield  {journal} {\bibinfo  {journal}
  {Science}\ }\textbf {\bibinfo {volume} {353}},\ \bibinfo {pages} {1526}
  (\bibinfo {year} {2016})}\BibitemShut {NoStop}%
\bibitem [{\citenamefont {Hwang}\ and\ \citenamefont
  {Sarma}(2008)}]{hwang2008}%
  \BibitemOpen
  \bibfield  {author} {\bibinfo {author} {\bibfnamefont {E.}~\bibnamefont
  {Hwang}}\ and\ \bibinfo {author} {\bibfnamefont {S.~D.}\ \bibnamefont
  {Sarma}},\ }\bibfield  {title} {\bibinfo {title} {Acoustic phonon scattering
  limited carrier mobility in two-dimensional extrinsic graphene},\ }\href@noop
  {} {\bibfield  {journal} {\bibinfo  {journal} {Physical Review B}\ }\textbf
  {\bibinfo {volume} {77}},\ \bibinfo {pages} {115449} (\bibinfo {year}
  {2008})}\BibitemShut {NoStop}%
\bibitem [{\citenamefont {Dunham}(1932)}]{dunham1932}%
  \BibitemOpen
  \bibfield  {author} {\bibinfo {author} {\bibfnamefont {J.}~\bibnamefont
  {Dunham}},\ }\bibfield  {title} {\bibinfo {title} {The
  wentzel-brillouin-kramers method of solving the wave equation},\ }\href@noop
  {} {\bibfield  {journal} {\bibinfo  {journal} {Physical Review}\ }\textbf
  {\bibinfo {volume} {41}},\ \bibinfo {pages} {713} (\bibinfo {year}
  {1932})}\BibitemShut {NoStop}%
\end{thebibliography}%
%Title of Sup Info
\pagebreak
\widetext
\begin{center}
\textbf{\large Supplementary Information}
\end{center}
\renewcommand{\thetable}{S\arabic{table}}
\renewcommand{\theequation}{S\arabic{equation}}
\renewcommand{\figurename}{Figure}
\renewcommand{\thefigure}{S\arabic{figure}}
\renewcommand\thesection{S\arabic{section}}
\setcounter{equation}{0}
\setcounter{figure}{0}
\setcounter{table}{0}
\setcounter{page}{1}
\makeatletter

\tableofcontents
\section{QPC conductances}
\begin{figure}[h]
    \centering
    \includegraphics[width=0.9\textwidth]{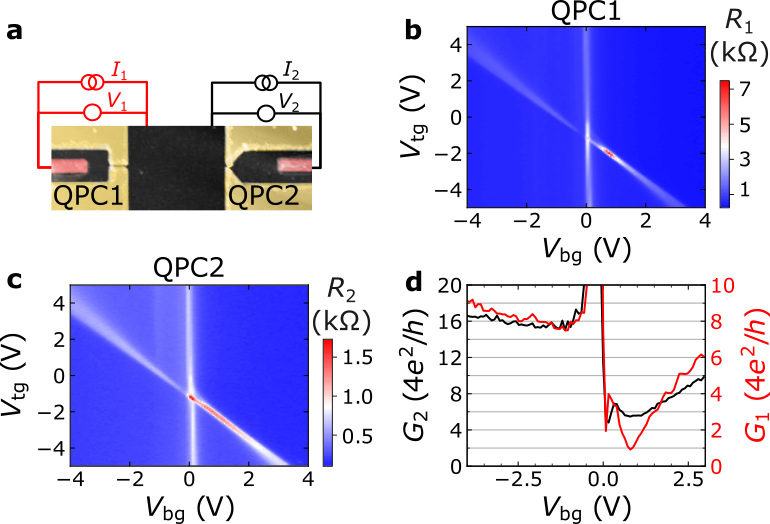}
    \caption{Conductance of QPC1 and QPC2. (a) Measurement geometry. The split gates are yellow, the contacts to BLG are red, and the measurement circuit is black. The split gate separation is 50~nm (b) and (c) Two-terminal resistance of QPC1 ($R_1$) and QPC2 ($R_2$), respectively as a function of \Vbg{} and \Vtg{} and measured using the red and black circuit in panel a, respectively.  (d) Conductance obtained from panels b and c as explained in the text.}
    \label{fig:FigS1}
\end{figure} 
To measure the conductance of the QPCs involved in the collimation experiments, we have used the two-terminal geometries shown in (Fig.~\ref{fig:FigS1}a). The red circuit is used for QPC1 and the black one for QPC2, where the applied current $I_1=I_2=100$~nA and the measured voltages are $V_1$ and $V_2$, respectively. In Fig.~\ref{fig:FigS1}b, we show the resistance map obtained from QPC1 ($R_1=V_1/I_1$) as a function of the back-gate voltage (\Vbg{}) and the top-gate voltage (\Vtg{}) applied to the corresponding split gates (yellow areas at the insets). The vertical line near \Vbg{}$=0$ corresponds to the charge neutrality point of the non-top-gated regions and the diagonal line to that of the top-gated ones. In Fig.~\ref{fig:FigS1}c, we show the resistance map obtained from QPC2 ($R_2=V_2/I_2$), showing similar features as QPC1 but a significantly ($\approx 4$ times) lower maximal resistance. 

To compare both QPCs, we have calculated the QPC conductances using $G(\Vbg{})=(R_\mathrm{max}(\Vbg{})-R_\mathrm{min}(\Vbg{}))^{-1}$, where $R_\mathrm{max(min)}(\Vbg{})$ is the maximal (minimal) resistance at each \Vbg{} ($R_\mathrm{min}(\Vbg{})$ is dominated by the contact resistances). The result is displayed in Fig.~\ref{fig:FigS1}d and, for QPC1, shows clear plateaus at $G=m\times4e^2/h$, where $m=$1, ..., 6. In contrast, QPC2 shows a smooth $G$ increase, indicating that the conductance is not quantized \cite{ingla2023}.
%\FloatBarrier
\section{\Vbg{}-dependence of the collimation spectra}
\begin{figure}[ht]
    \centering
    \includegraphics[width=0.85\textwidth]{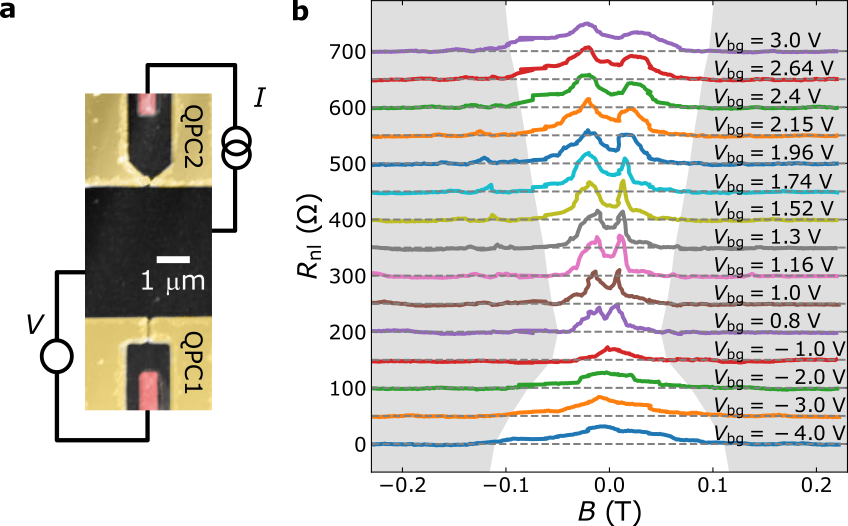}
    \caption{\Vbg{}-dependence of the collimation spectra. (a) Measurement geometry. The split gates are yellow, the contacts to BLG are red, and the measurement circuit is black. The QPC separation is 4~$\mu$m. (b) \Rnl{} vs.~$B$ for all the \Vbg{} used in Fig.~2 of the main manuscript. The curves are offset 500~$\Omega$ for clarity and the dashed lines show the offsets. The $B$ scale has been corrected for an offset corresponding to the magnet remanence.}
    \label{fig:FigS2}
\end{figure}
Here we show the complete \Vbg{}-dependence of the collimation signal described in Fig.~2 of the main manuscript.
The nonlocal signal, that was obtained using the configuration in Fig.~\ref{fig:FigS2}a, is defined using \Rnl{}$=V/I$, where $V$ is the measured voltage and $I$ is the applied current.
The result is plotted vs.~out-of-plane magnetic field ($B$) in Fig.~\ref{fig:FigS2}b, and shows two clear peaks for all \Vbg{}$>0$ and a single peak for \Vbg{}$<0$ with signatures of jet formation discussed in \ref{sectionTdep}. This result, together with the lack of size quantization for \Vbg{}$<0$, stresses the role of coherent injection at the QPC to achieve valley-polarized current jetting. Assuming a circular trajectory, one expects that \Rnl{}$>0$ only when the diameter of the cyclotron orbit ($2r_c=2\hbar\sqrt{n\pi}/(eB)$, where $\hbar$ is the reduced Plank's constant, $n$ the carrier density, and $e$ the electron charge) is greater than the QPC separation ($L=4\,\mu$m). For this reason, in Fig.~\ref{fig:FigS2}b, we have marked in gray the areas that do not fulfill this condition as a guide to the eye.
\begin{figure}
    \centering
    \includegraphics[width=\textwidth]{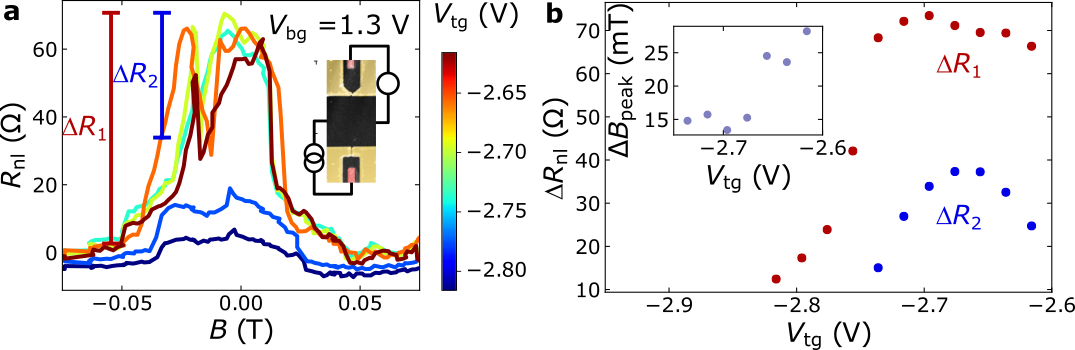}
    \caption{\Vtg{}-dependence of the collimation spectra. (a) \Rnl{} vs.~$B$ at \Vbg{}$=1.3$~V and \Vtg{} ranging from $-2.82$ to $-2.62$~V, as indicated by the color bar. The right inset shows the measurement geometry. The signal amplitudes $\Delta R_1$ and $\Delta R_2$ are shown at the right inset for \Vtg{}$\approx-2.7$~V. (b) $\Delta R_1$, $\Delta R_2$, and peak separation (\DBmax{}) vs.~\Vtg{} summarizing the effect of the electric field at the QPC on the collimation spectra.}
    \label{fig:FigS3}
\end{figure}
\FloatBarrier
\section{\Vtg{}-dependence of the collimation spectra}

Here we show collimation experiments obtained using the measurement circuit shown at the inset of Fig.~\ref{fig:FigS3}a. As above, the nonlocal signal is defined as $\Rnl{}=V/I$.
%The nonlocal measurements reported in the main manuscript are sensitive to the exact \Vtg{} applied to the split gates. The dependence is shown in Fig.~\ref{fig:FigS3}. 
In Fig.~\ref{fig:FigS3}, \Vbg{} is fixed to 1.3~V and \Rnl{} is measured as a function of $B$ at \Vtg{} ranging from -2.82 to -2.62~V. As shown in Fig.~\ref{fig:FigS3}b, \Vtg{} does not only influence the signal amplitude and the depth of the $B=0$ minimum but also the peak separations (\DBmax{}), which are plotted at the inset. This result shows that the actual shape of the confinement potential influences the jet formation. Such observation is consistent with the semi-classical model in \ref{sectionModel} and shows that the QPC plays a crucial role in the current jet formation. 

\FloatBarrier
\section{Temperature dependence}\label{sectionTdep}
\begin{figure}
    \centering
    \includegraphics[width=\textwidth]{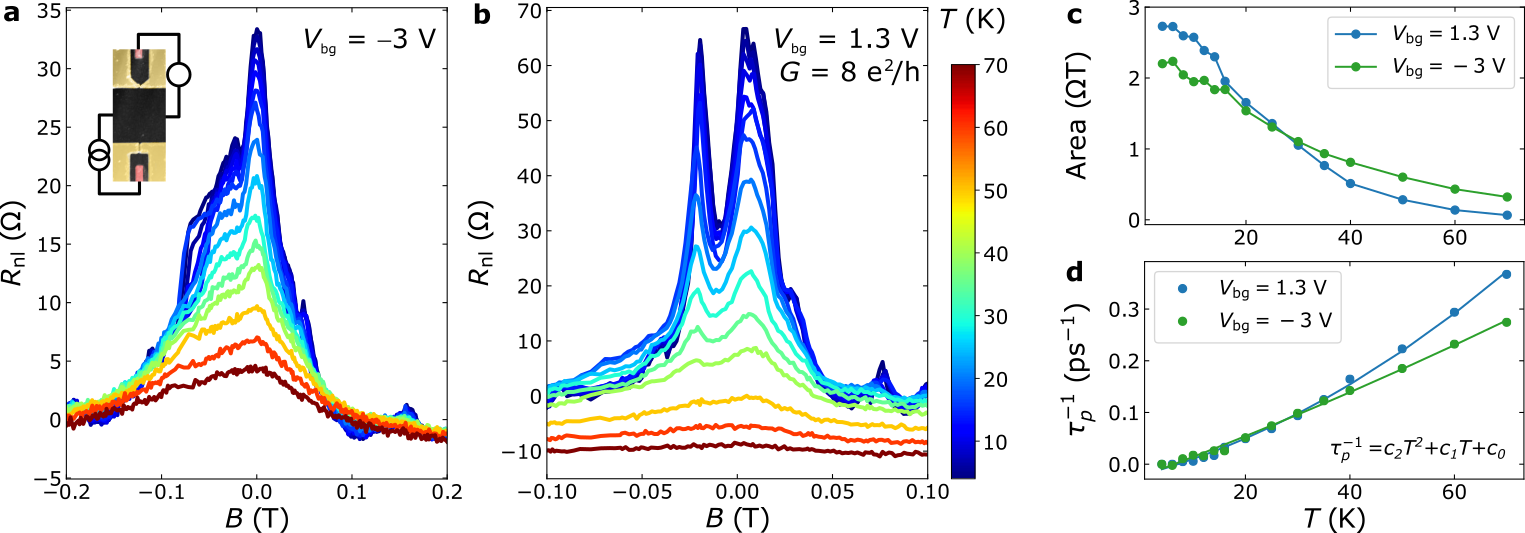}
    \caption{Temperature dependence of the collimation spectra. (a) \Rnl{} vs.~$B$ at \Vbg{}$=-3$~V (a) and \Vbg{}$=1.3$~V (b) and different temperatures, determined by the color scale in b. The measurement configuration is shown by the black circuit at the inset of panel a. (c) Area under the collimation spectra at different $T$ extracted from panels a and b. (d) Scattering rate obtained from the peak area dependence in panel c. The lines are fits to parabolas, as indicated at the inset.}
    \label{fig:FigS4}
\end{figure}

Here we show the temperature ($T$) dependence of the collimation signal at \Vbg{}$=-3$~V (Fig.~\ref{fig:FigS4}a) and \Vbg{}$=1.3$~V (Fig.~\ref{fig:FigS4}b). The former reveals that a second lobe is also present for negative \Vbg{}, even though the second peak observed for \Vbg{}$>0$ is not fully developed. As in the main manuscript, we argue that this is related to the absence of size quantization at QPC1 for \Vbg{}$<0$.
In both cases, the double-peak remains visible above 40~K, confirming that it is not a quantum interference effect \cite{kozikov2012}.   

The $T$-dependence of ballistic signals has been used to estimate the scattering rate ($\tau_p^{-1}$) \cite{yacoby1991,molenkamp1992,lee2016}. 
Here we use \cite{lee2016}
\begin{equation}
    \tau_p^{-1}(T)=-2 (v_F/L) \log(A(T)/A_0),
    \label{EqScattering}
\end{equation}
 where $v_F$ is the Fermi velocity, $L=4\,\mu$m is the QPC separation, $A(T)$ is the area under the peak at temperature $T$, and $A_0$ is the area at base temperature. The peak areas are shown in Fig.~\ref{fig:FigS4}c. Note that Equation~\ref{EqScattering} assumes that the ballistic electron trajectories are straight and their length is the QPC separation. Since the current jets are collected at a non-zero magnetic field, this is only an approximation and special care must be taken before drawing quantitative conclusions from this analysis.

In graphene, assuming that the QPCs define a hard-wall potential, a linear $T$-dependence of $\tau_p^{-1}$ is associated with electron-phonon scattering \cite{hwang2008}. In contrast, a quadratic $T$ dependence is associated with electron-electron interactions \cite{lee2016}. Simple inspection of the $\tau_p^{-1}$ vs.~$T$ plot in Fig.~\ref{fig:FigS4}d reveals that
the \Vbg{}$=1.3$~V data follow a more quadratic trend than the \Vbg{}$=-3$~V case, that looks more linear. This result is reflected on the fitting parameters in Table~\ref{tableTdep}, where the $T^2$ term ($c_2$) is 30 times larger at \Vbg{}$=1.3$~V, and indicates that electron-electron interactions may be more relevant in this case. Note that, even though some electron-hole asymmetry of the scattering rate has also been observed in Ref.~\cite{ingla2023} between \Vbg{}$=\pm3$~V, the smaller bandgap opening at \Vbg{}$=1.3$~V may influence the $T$-dependence and a definitive statement cannot be made. 
\begin{table}[]
    \centering
    \begin{tabular}{|c|c|c|c|}
        \hline
        \Vbg{} & $c_2$ &$c_1$ & $c_0$ \\
        (V) &(s$^{-1}$K$^{-2}$) & (s$^{-1}$K$^{-1}$) & (s$^{-1}$)\\
        \hline
         $1.3$ &$6\pm1\times 10^{-4}$&$-4\pm3\times 10^{-3}$&$7\pm16\times 10^{-3}$\\
         \hline
         $-3$ &$2\pm1\times 10^{-5}$&$1.19\pm0.09\times 10^{-2}$&$-8\pm1\times 10^{-2}$\\
         \hline
    \end{tabular}
    \caption{Parameters extracted from fitting $\tau_p^{-1}=c_2T^2+c_1T+c_0$ to the data in Fig.~\ref{fig:FigS4}d.}
    \label{tableTdep}
\end{table}
\FloatBarrier
\section{Reproducibility}
\begin{figure}
    \centering
    \includegraphics[width=0.65\textwidth]{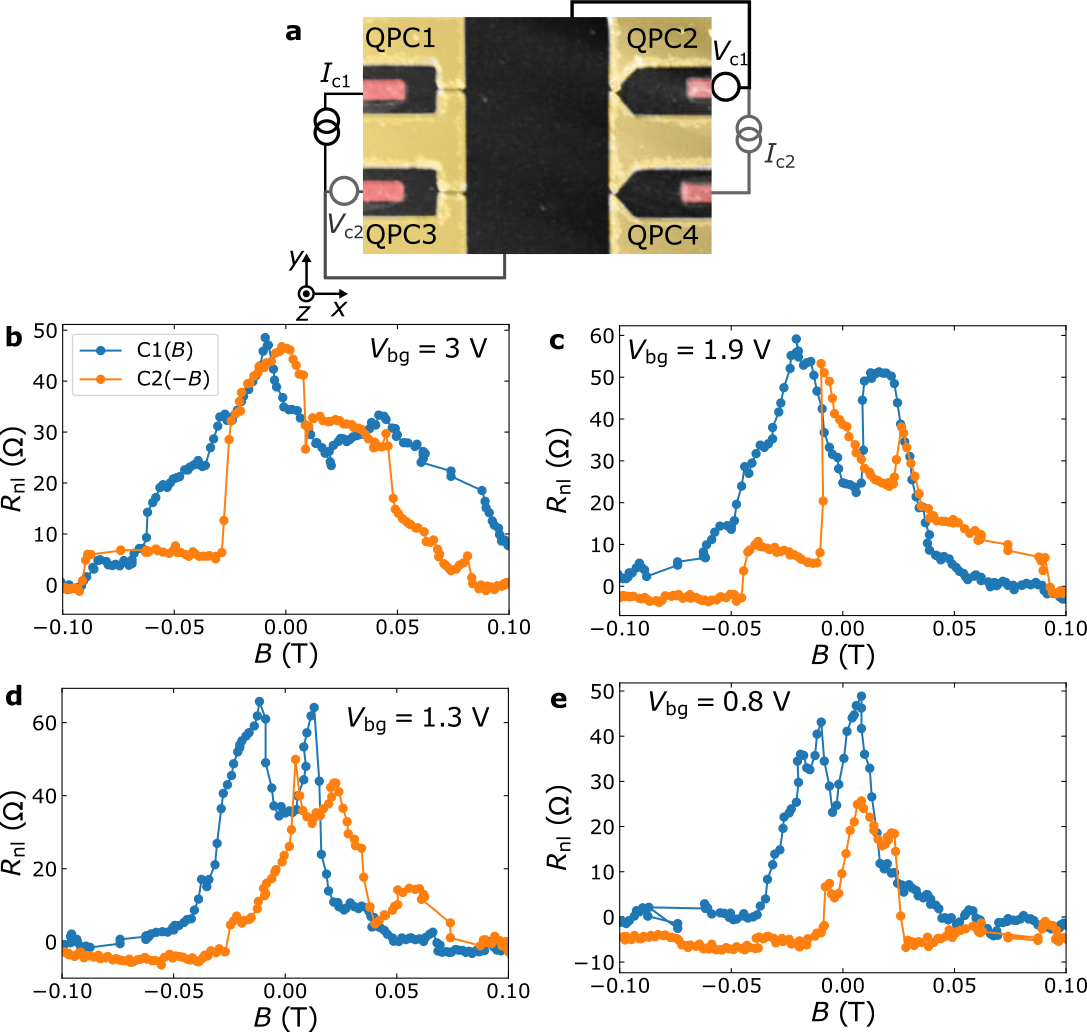}
    \caption{Reproducibility of the collimation spectra in a second QPC pair. (a) False-color atomic force microscope image of the measured device, corresponding to a 90$^\circ$ rotation of the inset of Fig.~\ref{fig:FigS2}a. The measurement configuration corresponding to C1 is shown by the black circuit and C2 by the gray circuit. \Rnl{} obtained from C1 and C2 are shown in blue and orange in panels (b)-(d) corresponding to \Vbg{}$=3$, $1.9$, $1.3$, and $0.8$~V, respectively.}
    \label{fig:FigS5}
\end{figure}
As mentioned in the main manuscript, the results obtained using the QPC pair formed by QPC1 and QPC2 have been reproduced in a second pair of QPCs (QPC3 and QPC4) which are aligned along the same crystallographic direction. The measurement configuration is shown in Fig.~\ref{fig:FigS5}a, where the source $I_{c1(2)}$ and voltage measurement unit $V_{c1(2)}$ are used for configuration C1(2). The experimental signals $\Rnl{}=V_{c1(2)}/I_{c1(2)}$ obtained for both geometries are shown in Figs.~\ref{fig:FigS5}b to \ref{fig:FigS5}e. Since the sources and measurement units are swapped between C1 and C2, we have reversed the $B$ for C2.
The \Vbg{}$=3$~V C2 data shows several sudden jumps and very broad peaks, indicating instabilities in the measurement and a slight miscalibration of \Vtg{} as shown in Fig.~\ref{fig:FigS3}. For this reason, the peak separation at \Vbg{}$=3$~V has not been included in Fig.~2e of the main manuscript. 

\FloatBarrier

\section{Semiclassical simulations}\label{sectionModel}

\subsection{Fermi surface calculation}

We consider the lowest electron-like band in bilayer graphene.
The dispersion in the vicinity of the $K$ and $K^{\prime}$ points is~\cite{mccann2013}
\begin{align}
    \varepsilon^2(p)&=\frac{\gamma_1^2}{2}+\frac{U^2}{4}+\left(v^2+\frac{v_3^2}{2}\right) p^2 - \sqrt{\Gamma} \\
    \Gamma&=\frac{1}{4}\left(\gamma_1^2-v_3^2 p^2\right)^2+v^2 p^2\left(\gamma_1^2+U^2+v_3^2 p^2\right) +2 \alpha \xi \gamma_1 v_3 v^2 p^3 \cos 3 \varphi,
\end{align}
where $v = \sqrt{3}a \gamma_0 / 2 \hbar$, $v_3 = \sqrt{3}a \gamma_3 / 2 \hbar$, $\gamma_0$ is the in-plane nearest-neighbors hopping strength, $\gamma_1$ is the nearest-neighbors out-of-plane hopping, and $\gamma_3$ is the next-nearest neighbors out-of-plane hopping; $U$ is the interlayer imbalance; $\xi=\pm 1$ is the valley number; $p$ is the momentum's radial component; and $\varphi$ is the momentum's polar angle.
We control the strength of the Fermi surface's trigonal warping with the parameter $\alpha$: when $\alpha=0$, there is no trigonal warping; whereas $\alpha=1$ corresponds to the fully warped surface.
We use the parameters from Ref.~\cite{kuzmenko2009}.

To compute the Fermi surface, we must first compute the chemical potential $\mu$, and the interlayer imbalance $U$ for a given set of electron density $n$ and external displacement field $D$.
First, we compute the screened interlayer imbalance~\cite{mccann2013}:
\begin{align}
    \frac{U(n)}{U_{\mathrm{ext}}} \approx & \left[1-\frac{\Lambda}{2} \ln \left(\frac{|n|}{2 n_{\perp}} + \frac{1}{2} \sqrt{\left(\frac{n}{n_{\perp}}\right)^2+\left(\frac{U}{2 \gamma_1}\right)^2}\right)\right]^{-1}
    \\
    n_{\perp} &= \frac{\gamma_1^2}{\pi \hbar^2 v^2}~, \quad \Lambda=\frac{c_0 e^2 \gamma_1}{2 \pi \hbar^2 v^2 \varepsilon_0 \varepsilon_r} \equiv \frac{c_0 e^2 n_{\perp}}{2 \gamma_1 \varepsilon_0 \varepsilon_r}
\end{align}
where $U_{\mathrm{ext}} = e d D$, and $d$ is the interlayer spacing.
We then compute the chemical potential $\mu$ by solving
\begin{align}
    n &= \int_{0}^{\mu} d\omega~A(U, \omega)~,
\end{align}
where $d$ is the interlayer spacing, and $A(U,\omega)$ is the density of states for the corresponding $U$.
Finally, we find the Fermi surface by solving $\varepsilon(\hbar k_F, \phi) = \mu$.

\subsection{Injection current}

Under a magnetic field $B$, the cyclotronic trajectories are parametrized as
\begin{equation}
    r(\phi) = \frac{\hbar}{a e B} k_F\left(\phi - \frac{\pi}{2}\right)~,
\end{equation}
where $a$ is the lattice constant, $\phi$ is the polar angle, $k_F=k_F(\phi)$ is the Fermi surface.
We neglect Berry curvature effects due to the large electron density in the experiment.

We consider electrons injected with an angle $\theta \in [0, \pi]$ from a point-like injector.
Thus, the efficiency of the injector at the injecting angle range $[\theta, \theta + d \theta]$ is
\begin{equation}
    \begin{aligned}
        dI(\theta) = \sum_n |\partial_{\phi}\theta| \cos\left(\theta(\phi) - \frac{\pi}{2}\right)  |\psi_n(\mathbf{k}(\phi))|^2 d\phi~.
    \end{aligned}
\end{equation}
where $\psi_m(\mathbf{k}(\theta))$ is the wavefunction of $m$-th mode in the QPC.

Since the collector showed no signs of quantization, we consider an isotropic \cite{molenkamp1990} efficiency corrected by the effect of trigonal warping
\begin{equation}
    dI(\theta_c, x_c) = |\partial_{\phi}\theta_c|\times\cos\left(\theta_c(\phi) - \frac{\pi}{2}\right) \eta\left(x_c(\phi)\right)~,
\end{equation}
where $\theta_c$ is the incidence angle at the collector, $x_c$ is the $x$-coordinate of the final point of the trajectory, and $\eta(x_c)$ is the efficiency of the QPC along the $x$-coordinate.
We assume a smooth-wall collector with
\begin{equation}
    \eta(x_c) = \frac{\left\lvert \mathrm{erf}\left(\frac{x_c + W / 2}{\sqrt{2} W}\right) - \mathrm{erf}\left(\frac{x_c - W / 2}{\sqrt{2}W}\right) \right\rvert}{2}~,
\end{equation}
where $\mathrm{erf}(x)$ is the error function.

\subsection{Size quantization effects}

The last step of the procedure is computing $|\psi_m(\mathbf{k}(\theta))|^2$, \emph{i.e.}, the occupation of the Fermi surface for each QPC mode.
We consider an infinitely long QPC with a parabolic confinement
\begin{equation}
    V(x) = \alpha x^2~,\quad \alpha = 4(E_{c,\mathrm{depleted}} - E_c)/W
\end{equation}
where $E_c$ and $E_{c,\mathrm{depleted}}$ are the bottom of the conduction bands at the QPC and in the depleted region, and $W$ is the QPC width.
Thus, the Hamiltonian in momentum space is
\begin{equation}
    \label{eq:momentum_hamiltonian}
    H(k_{\parallel}) = -\alpha\partial_{p_{\perp}}^2\psi + \varepsilon(p_{\perp}, \hbar k_{\parallel}) - \mu~,
\end{equation}
we used the QPC's translational symmetry to set $k_{\parallel}$ as a good quantum number.

For a given $\mu$, the propagating modes are zero-energy solutions of $H(k_{\parallel}) = H(k_{\parallel}^m)$, where $k_{\parallel}^m$ is the parallel momentum of the m-th mode.
We then estimate $k_{\parallel}^m$, and find the lowest-energy solution $\psi_m(\mathbf{k}(\theta))$ of $H(k_{\parallel}^m)$.
We estimate $k_{\parallel}^m$ using WKB approximation \cite{dunham1932}.
Namely, we solve
\begin{equation}
    \int_{\varepsilon(p_\perp, \hbar k_{\parallel}^m) - \mu < 0} dp~\sqrt{-\frac{\varepsilon(p_\perp, \hbar k_{\parallel}^m)}{\alpha}} = \left(m - \frac12 \right)\pi~.    
\end{equation}

Finally, we solve Eq.~\eqref{eq:momentum_hamiltonian} numerically to obtain $|\psi_m(\mathbf{k}(\theta))|^2$.

\subsection{Trigonal warping effect on the simulated peak separation}
\begin{figure}
    \centering
    \includegraphics[width=\textwidth]{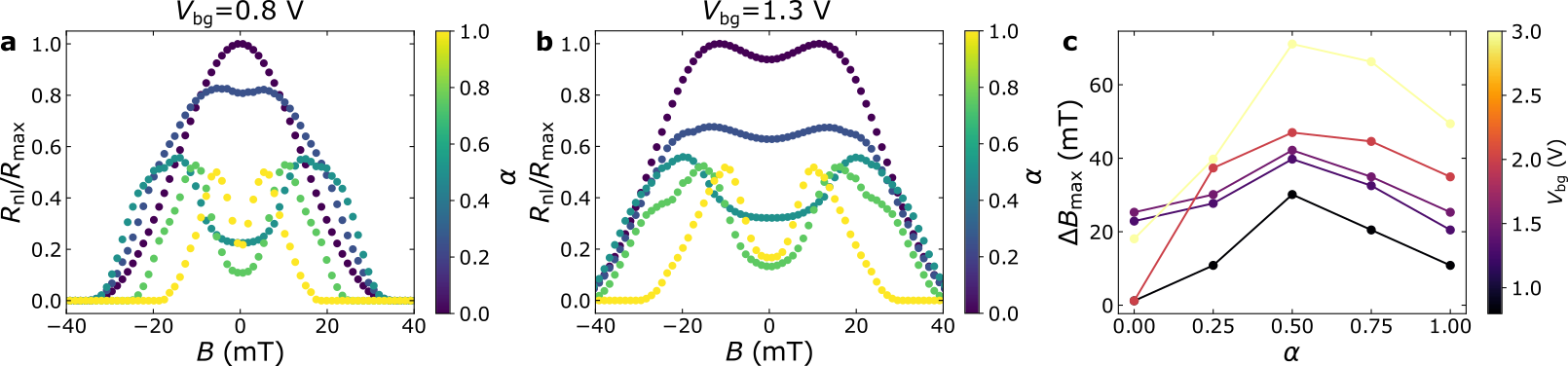}
    \caption{Effect of trigonal warping on the collimation spectra. Collimation spectra at $\Vbg{}=0.8$~V (a), corresponding to $m=1$, and $\Vbg{}=1.3$~V (b), corresponding to $m=2$ vs.~magnetic field. (c) Peak separation vs.~degree of trigonal warping for all the calculated \Vbg{}. }
    \label{fig:FigS7}
\end{figure}

We have simulated collimation spectra using the parameters in Table~\ref{table1}.
\begin{table}[]
    \centering
    \begin{tabular}{|c|c|c|c|c|c|c|c|c|}
        \hline
        $\gamma_0$ & $\gamma_1$ &$\gamma_3$ & $\gamma_4$  & $n$($\Vbg{}=3$ V)& $D$($\Vbg{}=3$ V)& $n$($\Vbg{}=0.8$ V)& $D$($\Vbg{}=0.8$ V) \\
        (eV) &(eV) & (eV) & (eV) & (m$^{-2}$)& (V/nm) & (m$^{-2}$)& (V/nm)\\
        \hline
         $3.16$ &$0.381$&$0.38$&$0.14$& $2.6\times 10^{16}$& $0.24$& $6.5\times 10^{15}$&0.059\\
         \hline
    \end{tabular}
    \caption{Parameters used to obtain the Fermi surfaces and simulated collimation spectra. The tight-binding parameters are obtained from Ref.~\cite{kuzmenko2009}, $n$ is obtained in Ref.~\cite{ingla2023} from the same sample using Shubnikov de Haas oscillations and $D = e/(2\epsilon_0)\alpha_\mathrm{bg}(\Vbg{}-\Vbg^0)$, where $\epsilon_0$ is the vacuum permittivity, $\alpha_\mathrm{bg}$ the back-gate lever arm and $\Vbg^0$ the charge neutrality point. Note that \Vtg{} is not included because these parameters describe the open BLG channel, not the depleted regions.}
    \label{table1}
\end{table}
As shown in the main manuscript, the separation between collimation peaks (\DBmax{}) is sensitive to the degree of trigonal warping of the Fermi surface. Here we represent the modeled dependence of \DBmax{} on $\alpha$ and describe the two regimes found. For small trigonal warping ($\alpha<0.5$), \DBmax{} increases with $\alpha$. In contrast, for $\alpha>0.5$ the opposite trend is observed. The evolution of the simulated spectra with $\alpha$ is shown in Figs.~\ref{fig:FigS7}a and \ref{fig:FigS7}b. In Fig.~\ref{fig:FigS7}a, for low $\alpha$, a minimum develops from the initial peak at $B=0$. The minimum is not fully developed until $\alpha = 0.5$ which corresponds to approximately the maximum peak separation. Further increasing $\alpha$ results in a decrease of the peak separation and narrowing of the peaks. In Fig.~\ref{fig:FigS7}b we illustrate the \Vbg{}$=1.3$~V case, which corresponds to $m=2$ in the QPC. The simulated spectra have two peaks separated from normal incidence even for $\alpha=0$, which is a consequence of the momentum-space distribution of the electron wave function at the QPC for $m=2$ \cite{topinka2000,shepard1992}. However, the simulated spectra show a minimum at $B=0$ which is not as deep as the experimental results. The experimental observation of two \Rnl{} peaks for $m=1$ to $6$ rules out this explanation as the source of the measured data.
The \DBmax{} vs.~$\alpha$ results are summarized for all \Vbg{} in Fig.~\ref{fig:FigS7}c.
\begin{figure}
    \centering
    \includegraphics[angle=270,origin=c,width=\textwidth]{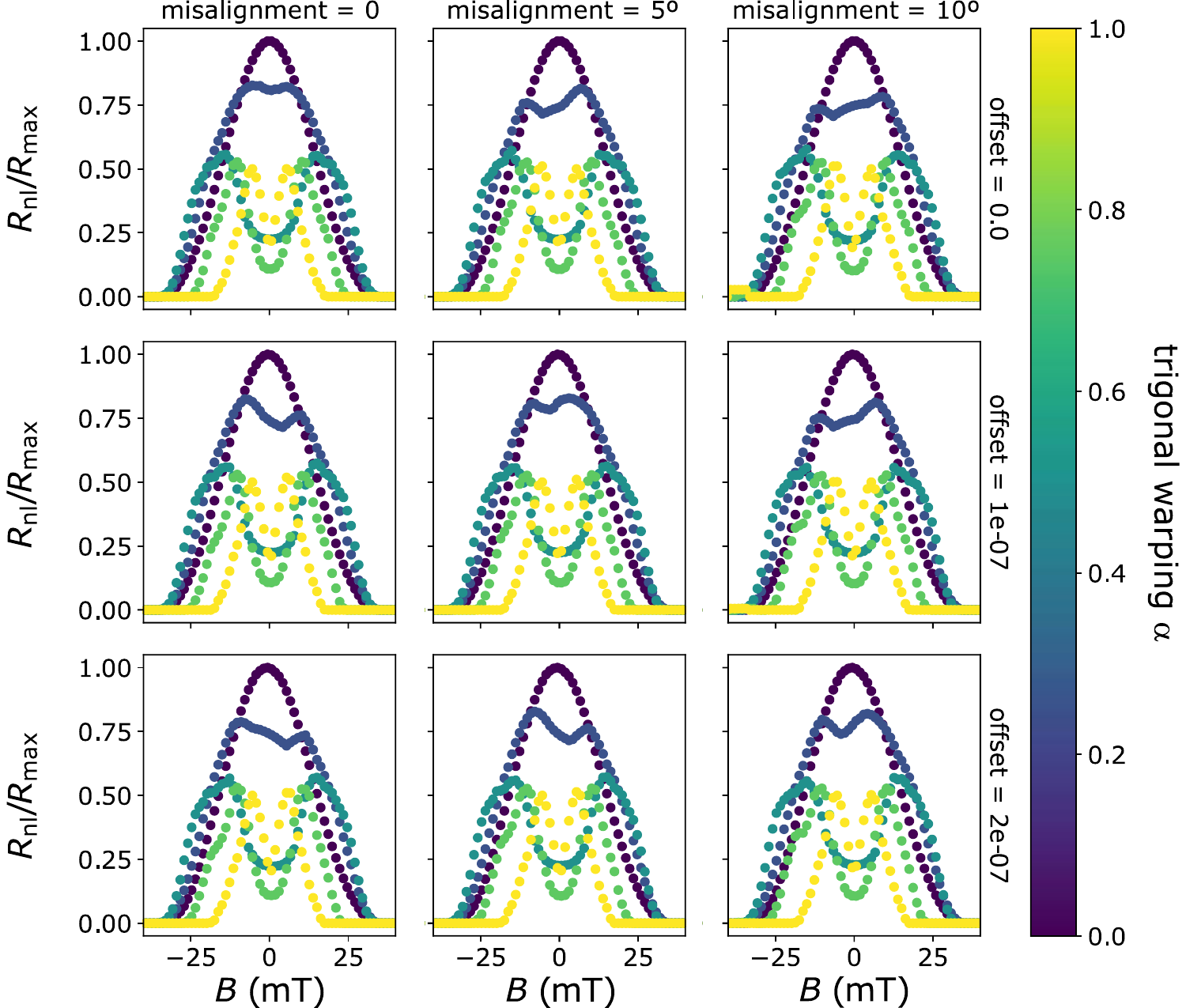}
    \caption{Simulated collimation spectra at \Vbg{}$=0.8$~V for $0^\circ$, $5^\circ$, and $10^\circ$ misaligned QPCs with respect to the BLG crystal (columns) and 0, 100, and 200~nm offset detector (rows). Each plot shows five different degrees of trigonal warping.}
    \label{fig:FigS6}
\end{figure}
\subsection{Effect of tilt and misalignment on the simulated collimation spectra}

The simulations shown until this point assume perfect alignment of the QPC split gates with a crystallographic direction of the BLG and injector and collector QPCs that are perfectly opposing each other. However, real devices may suffer from misalignments between the gates and the BLG crystal, and small distortions of the top gate pattern may result in a small offset along $y$ (see Fig.~\ref{fig:FigS5}a) of the detector position with respect to the injector, leading to non-perfectly opposing QPCs. We have introduced these features in the simulated spectra shown in Fig.~\ref{fig:FigS6}. As shown here, even though a misalignment of $10^\circ$ alone does not give rise to a significant asymmetry, together with a detector offset of 200~nm it gives rise to a sizeable asymmetry between the width of the positive and negative $B$ peak while the peak heights remain similar. We thus attribute the observed asymmetry to the combination of a misalignment with the BLG crystal and a small shift of the detector with respect to the injector. 
\FloatBarrier
%\bibliography{bibliography}
\end{document}